\documentclass[12pt,a4paper]{article}
\usepackage{authblk}
\usepackage{natbib}
\usepackage{amssymb}                    
\usepackage{amsmath}
\usepackage{chngcntr}  
\usepackage{booktabs}  
\usepackage{setspace}
\usepackage{float}
\usepackage{a4wide}
\usepackage{lineno} 
\usepackage[symbol]{footmisc}
\usepackage{lipsum}

\usepackage{graphicx}
\usepackage{color}

\renewcommand{\thefootnote}{\fnsymbol{footnote}}

\title{Fully dynamic earthquake cycle simulations on a non-planar fault using the spectral boundary integral element method.}

\author[1*]{Pierre ROMANET}
 \author[2]{So OZAWA}
\affil[1]{Earthquake and Tsunami research division, NIED, Tsukuba, Japan, email: romanet@bosai.go.jp}
\affil[2]{Earth and Planetary science, School of Science, The University of Tokyo, Tokyo, Japan, email: sozawa@eps.s.u-tokyo.ac.jp}
\date{}

\begin{document}
\maketitle
\doublespacing
\subsubsection*{Declaration of Competing Interests}
{\let\thefootnote\relax\footnote{{* Corresponding author: ROMANET Pierre,
Earthquake and Tsunami Research Division,
National  Research Institute for Earth Science and Disaster Resilience,
3-1, Tennodai, Tsukuba, Ibaraki, 305-0006, JAPAN}}}
The authors acknowledge there are no conflicts of interest recorded.
\newpage
\begin{abstract}
One of the most suitable methods for modeling fully dynamic earthquake cycle simulations is the spectral boundary integral element method (sBIEM), which takes advantage of the fast Fourier transform (FFT) to make a complex numerical dynamic rupture tractable. However, this method has the serious drawback of requiring a flat fault geometry due to the FFT approach. Here we present an analytical formulation that extends the sBIEM to a mildly non-planar fault. We start from a regularized boundary element method and apply a small-slope approximation of the fault geometry. Making this assumption, it is possible to show that the main effect of non-planar fault geometry is to change the normal traction along the fault, which is controlled by the local curvature along the fault. We then convert this space--time boundary integral equation of the normal traction into a spectral-time formulation and incorporate this change in normal traction into the existing sBIEM methodology. This approach allows us to model fully dynamic seismic cycle simulations on non-planar faults in a particularly efficient way. We then test this method against a regular boundary integral element method for both rough-fault and seamount fault geometries, and demonstrate that this sBIEM maintains the scaling between the fault geometry and slip distribution. 
\end{abstract}

\newpage
\section{Introduction} 
The effect of fault geometry on earthquake mechanics is one of the most fundamental questions in seismology. Numerous numerical \citep{aochi2000b,bhat2004,oglesby2005,dunham2011b,mitsui2018,ando2018,wollherr2019,sathiakumar2021} and theoretical \citep{poliakov2002,rice2005,fang2013} studies have made significant advances in improving our understanding of the effect of complex  fault geometries and their associated seismogenesis. However, a major limiting factor in the dynamic rupture modeling of complex fault geometries and/or modeling in a complex medium is the computational time required to run such a complex earthquake model. 

The boundary integral element method (BIEM) is one approach for obtaining an analytical solution of the stress field in the form of an integral over the history of the slip distribution along the fault. The analytical nature of the BIEM means that it has strong limitations, such as the requirement of a simple medium during simulations. The elasto-dynamic BIEM involves a space--time convolution when modeling planar faults, such that the simulation can be accelerated by calculating the stress in the wavenumber domain via the Fast Fourier Transform (FFT). This appears to have been used for the first time in earthquake mechanics by \citet{andrews1985}, where a Fourier transform was employed over the Green’s function that was discretized in time. \citet{perrin1995} was the first to employ the Fourier transform on the continuous BIEM to find a representation that links the slip and stress on the fault in the wavenumber domain. This method was quickly generalized to other modes and three-dimensional (3D) problems by \citet{geubelle1995}. It has since been greatly improved by \citet{lapusta2000} to run multi-cycle simulations of dynamic rupture and aseismic slip, with a particular improvement being the ability to determine a very efficient way to perform a wavenumber-dependent truncation of the integral. This method is an extremely efficient method for performing numerical simulations of earthquake cycles, and was generalized to simulate a 3D model space in the late 2000s \citep{lapusta2009a}. We will hereafter call this BIEM that uses FFTs the spectral BIEM (sBIEM). This paper will refer only to fully dynamic simulations, although spectral methods also exist for static and quasi-dynamic simulations.
One drawback of the sBIEM is that it is not applicable to non-planar faults due to the requirement of equi-spaced data when using FFT. Therefore, fully dynamic earthquake cycle simulations on non-planar faults require much larger computational costs than those with planar faults. A literature search has indicated that a fully dynamic earthquake cycle simulation on non-planar faults that employs space--time BIEM has not been conducted to date, which is most likely due to the above-mentioned computational constraint. Although the quasi-dynamic approach is a popular approach in earthquake cycle simulations \citep{rice1993,romanet2018,ozawa2020}, it simplifies the wave-mediated stress transfer and yields large differences compared with fully dynamic simulations \citep{lapusta2009a,thomas2014c}. Therefore, more efficient ways to simulate the earthquake cycle via a fully dynamic BIEM that is applicable to non-planar faults are needed. 

One way would be to extend the applicability of the sBIEM to non-planar geometries, as mentioned by \citet{heimisson2020}, who suggested that including roughness drag \citep{fang2013} in the sBIEM may allow the ability to account for geometrical effects. Another way would be to introduce normal traction variations along the fault. This approach has been employed successfully in simulating the geometrical variations of a seamount \citep{yang2013} using a classic space--time BIEM; it has since been simulated using sBIEM \citep{schaal2019}. This later approach can be justified via the observation that the main effect of non-planar fault geometry is to change the normal traction along the fault \citep{chester2000,tal2018}, which has been recently demonstrated analytically in the static case \citep{romanet2020, cattania2021}. However, this approach is currently a crude approximation because the normal traction must vary with on-going slip and cannot be chosen randomly. Finally, we note that the sBIEM has been generalized recently to simulate parallel faults in a quasi-dynamic model \citep{barbot2021}.

Here we present an analytical approach that allows us to extend the fully dynamic sBIEM to a non-planar fault geometry. We take advantage of a recently developed planar fault approximation (the small-slope approximation) that retains the zeroth-order effect on the stress for a non-planar geometry while keeping the fault planar \citep{romanet2020}. We only apply this method to an in-plane pure shear fault (mode II), although it can be generalized to other mode of slip (mode I, opening faults). The out-of-plane pure shear case (mode III) corresponds to the case where there is no curvature along the slip direction, such that there is no zeroth-order effect of the geometry on the stress. 
 
 %

\section{Boundary element equations for the small-slope approximation and its spectral representation}
\subsection{The regularized boundary element equation}
A fully dynamic regularized boundary element method for an in-plane pure shear (mode II) fault that is embedded in a two-dimensional (2D) linear elastic medium can be written as \citep{bonnet1999,sato2020,romanet2020}:
\begin{equation}
\begin{split}
&\sigma_{ab}^{\text{el}}(\mathbf{x},t) = \\
&\text{\underline{Gradient of slip:}}\\
&c_{abcd}\int_{0}^{t}  \iint _{\Sigma}c_{ijpq}\frac{\partial}{\partial x_q} G_{cp} (\mathbf{x}-\mathbf{y},t-\tau) [n_d (\mathbf{y}) t_j (\mathbf{y}) - n_j(\mathbf{y})  t_d(\mathbf{y}) ] t_i (\mathbf{y}) \left[ \frac{\partial}{\partial y^t} \Delta u^t(\mathbf{y} ,\tau)\right]   \mathrm{d}\Sigma \mathrm{d}\tau \\
&\text{\underline{Curvature $\times$ Slip:}}\\
+&c_{abcd}\int_{0}^{t}  \iint _{\Sigma}c_{ijpq}\frac{\partial}{\partial x_q} G_{cp}(\mathbf{x}-\mathbf{y},t-\tau)[n_d(\mathbf{y})  t_j(\mathbf{y}) - n_j(\mathbf{y})  t_d(\mathbf{y}) ]  n_i (\mathbf{y})  \left[ \kappa^t(\mathbf{y})   \Delta u^t(\mathbf{y},\tau)  \right]   \mathrm{d}\Sigma \mathrm{d}\tau  \\
&\text{\underline{Inertia term:}}\\
-&\rho c_{abcd}\int_{0}^{t}\iint _{\Sigma}    \frac{\partial^2}{\partial t^2}G_{ic}(\mathbf{x}-\mathbf{y},t-\tau) n_d (\mathbf{y}) \Delta u^t(\mathbf{y},\tau) t_i \mathrm{d}\Sigma  \mathrm{d}\tau,
\label{main_eq}
\end{split}
\end{equation}
where $\sigma_{ab}^{\text{el}}(\mathbf{x},t)$ is the $ab$ component of the elastic stress due to a slip distribution at point $\mathbf{x}= \{x_1,x_2\}$ and time $t$, $c_{ijpq}$ is the $ijpq$ component of the Hooke tensor, $n_d$ and $t_j$ are the $d$ and $j$ components of the normal $\mathbf{n}$ and tangential vector $\mathbf{t}$ along the fault, respectively, $G_{cp}$ is the $cp$ component of the Green’s function for an infinite homogeneous medium, $\frac{\partial}{\partial y^t} \Delta u^t = \left(t_1\frac{\partial}{\partial x_1}+t_2\frac{\partial}{\partial x_2}\right)\Delta u^t$ is the derivative of the shear slip, $\kappa^t \Delta u^t$ is the product of the curvature $\kappa^t $ and shear slip $\Delta u^t$ along the fault, and $\rho$ is the density of the homogeneous medium. Please note that the sub-indices represent the components of a given vector in the global coordinate system (e.g., $x_q$ is the $q$ component of the $\mathbf{x}$ vector). Furthermore, the integrations are performed over a fault that is represented by $\Sigma$, where $\mathbf{y}=\{y_1,y_2\}$ is a vector that belongs to $\Sigma$.

The 2D dynamic Green's functions at position $\mathbf{x}$ and time $t$ from a source point at position $\mathbf{y}$ and time $\tau$ are given by \citet{tada1997}: 
 \begin{equation}
\begin{split}
G_{ij}(x,y,t,\tau)&=\frac{c_s^2}{2 \pi  \mu r^2 \sqrt{(t-\tau)^2-(r/c_p)^2)}}\left(\gamma_i \gamma_j \left[ 2(t-\tau)^2-\frac{r^2}{c_p^2} \right] -\delta_{ij}  \left[ (t-\tau)^2-\frac{r^2}{c_p^2} \right]  \right) \mathcal{H}\left( t-\tau-\frac{r}{c_p} \right)\\
&-\frac{c_s^2}{2 \pi  \mu r^2 \sqrt{2(t-\tau)^2-(r/c_s)^2)}}\left(\gamma_i \gamma_j \left[ 2(t-\tau)^2-\frac{r^2}{c_s^2} \right]- \delta_{ij}   (t-\tau)^2  \right) \mathcal{H}\left( t-\tau-\frac{r}{c_s} \right),
\end{split}
\label{green1}
\end{equation}
where:
\begin{equation}
\begin{split}
r(\mathbf{x},\mathbf{y}) &= \sqrt{(x_1-y_1)^2+(x_2-y_2)^2}, \\
\gamma_1(\mathbf{x},\mathbf{y}) &= \frac{(x_1-y_1)}{r(\mathbf{x},\mathbf{y})}, \\
\gamma_2(\mathbf{x},\mathbf{y}) &= \frac{(x_2-y_2)}{r(\mathbf{x},\mathbf{y})},
\end{split}
\end{equation}
$c_p$ and $c_s$ are the compressional and shear wave speeds, respectively, $\mu$ is the shear modulus, and $\delta_{ij}$ and $\mathcal{H}$ are the Kronecker and Heaviside functions, respectively.

\subsection{The small-slope approximation}
\begin{figure}[H]
\centering
\includegraphics[width=\textwidth]{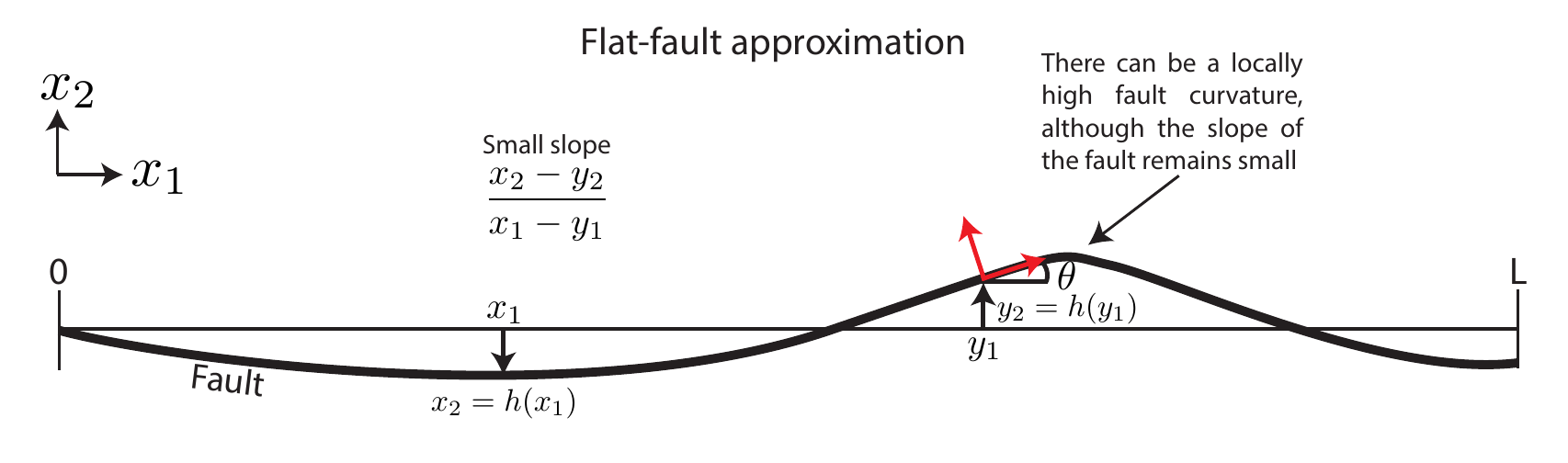}
\caption{The flat-fault approximation (modified from \cite{romanet2020}, fig. 3). It is possible to simplify the BIEM for the traction along a given fault if the slope between any pair of points along the fault is small. }
\label{flat_fault}
\end{figure}
The motivation for the small-slope approximation derived in \citet{romanet2020} comes from the approximations derived by \citet{saucier1992}, \citet{chester2000}, \citet{dunham2011b}, and \citet{fang2013} for linear perturbation analysis of the fault geometry. Here we consider that the fault slope is sufficiently small to neglect some terms in the BIEM. The stress on the fault and in the medium generally depends on both the slip gradient along the fault and the local curvature of the fault, which is multiplied by the slip along the fault (eq. \eqref{main_eq}). 
When we applied the small-slope approximation in the static state, we demonstrated in \citet{romanet2020} that the zeroth-order effect of the normal and shear tractions along a given fault did not depend on both the curvature and gradient terms but only on one of these terms. The normal traction depends mainly on the fault curvature, which is multiplied by the slip along the fault, while the shear traction depends mainly on the slip gradient along the fault. 
Previous results and the fact that the Coulomb friction links the shear traction $\tau$ to the normal traction $\sigma_n$ via the friction coefficient $f$ ($\tau=f\sigma_n$) made it possible to obtain a scaling relationship that linked the fault geometry and the shear slip distribution: $\frac{d \Delta u^t}{ \Delta u^t} \propto -f dm $. This scaling means that the relative variation in shear slip $\frac{d \Delta u^t}{\Delta u^t}$ is proportional to the product of the variation in slope along the fault $dm$ and the friction coefficient $f$ (please note that there was a sign mistake in \citet{romanet2020}, coming the expression of the kernels). One limitation of this scaling is that it does not apply near the edge of the fault since the slip gradients are usually very high in this region. 

We reiterate that each receiver point $\mathbf{x}= \{x_1,x_2\}$ will be a point where the stress and/or displacement is calculated, and that each source point $\mathbf{y}=\{y_1,y_2\}$ will be a point that creates the interaction with point $\mathbf{x}$ (this is a source point).
We will hereafter make the assumption that the fault lies in the $x_2=0$ plane, as this can be easily generalized to any kind of planar fault. The flat-fault approximation is the zeroth-order approximation that results from the assumption that the slope between any two points along the fault remains small ($\frac{x_2-y_2}{x_1-y_1}<<1$). Some of the following simplifications arise:
\begin{equation}
\begin{split}
r(\mathbf{x},\mathbf{y}) &\simeq \sqrt{(x_1-y_1)^2} ,\\
\gamma_1(\mathbf{x},\mathbf{y}) &\simeq \mathrm{sgn}(x_1-y_1), \\
\gamma_2(\mathbf{x},\mathbf{y}) &\simeq 0 ,\\
\end{split}
\end{equation}
where $\mathrm{sgn}$ is the sign function.
Another simplification comes for the tangential and normal vectors along the fault due to the small-slope assumption:
\begin{equation}
\begin{split}
\mathbf{t} &= \{ \cos (\theta) , \sin (\theta) \} \simeq  \{1, 0\},\\
\mathbf{n} &=  \{ -\sin (\theta) , \cos(\theta) \} \simeq \{0, 1\},
\end{split}
\end{equation}
where $\theta$ is the angle between the horizontal axis $\mathbf{e}_1$ and the tangent to the fault.
We can obtain the elastic tangential traction $\tau^{\text{el}}$ and the elastic normal traction $\sigma_n^{\text{el}}$ based on the previous equations and by retaining only the zeroth-order effect as follows: 
\begin{equation}
\begin{split}
\tau^{\text{el}} &=\mathbf{t}\cdot \overline{\overline{\sigma^{\text{el}}}}\cdot \mathbf{n} \simeq   \sigma_{12}^{\text{el}},  \\
\sigma_n^{\text{el}} &=\mathbf{t}\cdot \overline{\overline{\sigma^{\text{el}}}}\cdot \mathbf{n} \simeq  \sigma_{22}^{\text{el}}.
\end{split}
\end{equation}

\subsection{Spectral formulation for shear traction}
We employed the assistance of Mathematica to fully develop the boundary element method (eq. \eqref{main_eq}), together with the formulation of the Green's function (eq. \eqref{green1}), and then apply the small-slope assumption to obtain an expression for shear traction $\tau^{\text{el}}$:
\begin{equation}
\begin{split}
&\tau^{\text{el}}(\mathbf{x},t) = \\
-&\int_{0}^{t}  \int _{\Sigma} \frac{\partial}{\partial  t} \left[ \frac{2c_s^2\mu}{\pi c_p^3(t-\tau)} \frac{c_p^3(t-\tau)^3}{(x_1-y_1)^3}\sqrt{1-\frac{(x_1-y_1)^2}{c_p^2(t-\tau)^2}}\right] \mathcal{H}\left( t-\tau-\frac{ | x_1-y_1|}{c_p} \right) \\
& \qquad \qquad  \left[   \frac{\partial}{\partial \xi} \Delta u^t  \right]   d\Sigma d\tau  +\\
+&\int_{0}^{t}  \int _{\Sigma} \frac{\partial}{\partial t} \left[  \frac{2\mu}{\pi c_s(t-\tau)}\frac{c_s^3(t-\tau)^3}{(x_1-y_1)^3}\sqrt{1-\frac{(x_1-y_1)^2}{c_s^2(x_1-y_1)^2}} \right] \mathcal{H}\left( t-\tau-\frac{ | x_1-y_1|}{c_s} \right) \\
& \qquad \qquad  \left[  \frac{\partial}{\partial \xi} \Delta u^t \right]   d\Sigma d\tau  \\
+&\int_{0}^{t}  \int _{\Sigma} \frac{\partial}{\partial t} \left[  \frac{\mu}{2\pi c_s(t-\tau)}\frac{c_s(t-\tau)}{(x_1-y_1)}\frac{1}{\sqrt{1-\frac{(x_1-y_1)^2}{c_s^2(t-\tau)^2}}} \right] \mathcal{H}\left( t-\tau-\frac{ | x_1-y_1|}{c_s} \right) \\
& \qquad \qquad  \left[  \frac{\partial}{\partial \xi} \Delta u^t \right]   d\Sigma d\tau  \\
-&\int_{0}^{t}  \int _{\Sigma}  \frac{\mu}{2\pi c_s^2}\frac{\Delta u^t}{\left((t-\tau)^2-\frac{(x_1-y_1)^2}{c_s^2}\right)^{3/2}} \mathcal{H}\left( t-\tau-\frac{ | x_1-y_1|}{c_s} \right)   d\Sigma d\tau.
\end{split}
\label{shear}
\end{equation}
The shear traction that is determined in eq. \eqref{shear} is exactly the shear traction that one would find for a flat fault. Note that the last term, which is the inertia term for S waves, is the same as eq. 6 of \citet{cochard1994}; therefore, this term contains the radiation damping term $-\frac{\mu V}{2c_s}$ \citep{rice1993}. The radiation damping term accounts for the instantaneous response of the shear traction to slip velocity. We observe that the first-order perturbation of the fault slope does not change the zeroth-order shear traction. This result, which was obtained via a fully dynamic BIEM, is similar to the static result \citep{romanet2020}, whereby the zeroth-order shear traction depends mainly on the slip gradient along the fault.

Here we employ the spectral formulation that has been developed by \citet{geubelle1995} for the flat-fault case:
\begin{equation}
\tau^{\text{el}}(k,t) = -\mu |k| \int_0^t c^T_{II}(k,\tau)\Delta u^t(k,t-\tau)d\tau,
\end{equation}
where : 
\begin{equation}
 c_{II}^T(k,x)=\frac{1}{2}\left( \frac{J_1(c_s k x)}{ x}+4c_s k^2 x^2[W(c_p k x)-W(c_s k x)]-4\frac{c_s^2}{c_p} kJ_0(c_p k x )+3c_s k J_0(c_s k x)\right),
 \end{equation}

$J_0$ and $J_1$ are the zeroth- and first-order Bessel functions, respectively. The function W is simply the opposite of the Bessel integral
function of the first-order $J_{i_1}$ \citep{humbert1933}:
\begin{equation}
W(x)= -J_{i_1} (x),
\end{equation}
The function $W$ can be written with the help of usual functions as: 
\begin{equation}
W(x) =1- xJ_0(x)+J_1(x)-\psi(x),
\end{equation}
where $\psi$ is defined by the following expression: 
\begin{equation}
\psi(x) = \frac{\pi}{2}x(J_1(x)H_0(x)-J_0(x)H_1(x)),
\label{psi}
\end{equation}
and $H_0$ and $H_1$ are the zeroth- and first-order Struve functions, respectively.

 Integration by parts yields an equation that is more suitable for integration by changing the slip $\Delta u^t$ to the slip rate $\Delta \dot{u}^t$ and explicitly separating the static and dynamics terms (eq 5 in \citet{lapusta2009a}): 
\begin{equation}
\tau^{\text{el}}(k,t) = -\underbrace{\mu |k| \left(1-\frac{c_s^2}{c_p^2} \right) \Delta u^t(k,t)}_{\text{Static term}} - \underbrace{\mu |k| \int_0^t C^T_{II}(k,\tau) \Delta \dot{u}^t(k,t-\tau)d\tau}_{\text{Dynamic term}},
\end{equation}
where:
\begin{equation}
\begin{split}
C^T_{II}(t)&= \int c_{II}^T(k,t)dt\\
&=  \left(c_s^2 k^2t^2+\frac{c_s^2}{c_p^2}\right)W(c_p k t)-\left(c_s^2 k^2t^2+1\right)W(c_s k t) \\
    &-\frac{c_s^2}{c_p^2}J_1(c_p k t)+\frac{1}{2}J_1(c_s k t)\\
    &-\frac{c_s^2}{c_p}ktJ_0(c_p k t)+c_s k tJ_0(c_s k t).
    \end{split}
    \label{Ct}
\end{equation} 
 $C^T_{II}(t)$ seems to have been integrated numerically in \citet{lapusta2009a}. We are providing an analytical expression for $C^T_{II}(t)$ here (eq. \eqref{Ct}; further details on the calculation are available in Appendix \ref{calculationT}; see also \citet{noda2021}).
We note that the static and dynamic terms cancel each other at $t=0$ when $W(0)=1$, $J_1(0)=0$, and $J_0(0)=1$.

\subsection{Spectral formulation for normal traction}
We have also fully developed the regularized boundary integral equation (eq. \eqref{main_eq}) for normal traction $\sigma_n^{\text{el}}$ using Mathematica, and then applied the small-slope approximation:
\begin{equation}
\begin{split}
&\sigma_n^{\text{el}}(\mathbf{x},t) = \\
&\int_{0}^{t}  \int _{\Sigma} \frac{\partial}{\partial  t} \left[  \frac{4c_s^2 \mu(t-\tau)}{2\pi(x_1-y_1)^3}\sqrt{(t-\tau)^2-\frac{(x_1-y_1)^2}{c_p^2}}\right] \mathcal{H}\left( t-\tau-\frac{ | x_1-y_1|}{c_p} \right)\left[  \kappa^t (\mathbf{y})  \Delta u^t(\mathbf{y},\tau) \right]   d\Sigma d\tau  \\
+&\int_{0}^{t}  \int _{\Sigma} \frac{\partial}{\partial  t} \left[ \frac{c_s^2\lambda^2(t-\tau)}{2\pi c_p^2(x_1-y_1)\sqrt{(t-\tau)^2-\frac{(x_1-y_1)^2}{c_p^2}}} \right] \mathcal{H}\left( t-\tau-\frac{ | x_1-y_1|}{c_p} \right)\left[  \kappa^t (\mathbf{y})   \Delta u^t(\mathbf{y},\tau) \right]   d\Sigma d\tau  \\
-&\int_{0}^{t}  \int _{\Sigma} \frac{\partial}{\partial t} \left[  \frac{4c_s^2 \mu(t-\tau)}{2\pi(x_1-y_1)^3}\sqrt{(t-\tau)^2-\frac{(x_1-y_1)^2}{c_s^2}} \right] \mathcal{H}\left( t-\tau-\frac{ | x_1-y_1|}{c_s} \right) \left[  \kappa^t (\mathbf{y})   \Delta u^t(\mathbf{y},\tau) \right]   d\Sigma d\tau  \\
+&\int_{0}^{t}  \int _{\Sigma} \frac{\partial}{\partial t} \left[  \frac{c_s^2(\lambda+2\mu)(t-\tau)}{2\pi c_s^2(x_1-y_1)\sqrt{(t-\tau)^2-\frac{(x_1-y_1)^2}{c_s^2}}}  \right] \mathcal{H}\left( t-\tau-\frac{ | x_1-y_1|}{c_s} \right) \left[  \kappa^t (\mathbf{y})   \Delta u^t(\mathbf{y},\tau)  \right]   d\Sigma d\tau,  
\end{split}
\end{equation}
where $\lambda$ is the second Lam\'{e} parameter. It is possible to write the normal traction in the wavenumber domain in a similar way to the shear traction (see Appendix \ref{calculationN} for the full calculation):

\begin{equation}
\begin{split}
\sigma_n^{\text{el}}(k,t) =  i \underbrace{\mu  \left(1-\frac{c_s^2}{c_p^2} \right) \mathcal{F}[\kappa^t(x_1)\Delta u^t(x_1,t)]}_{\text{Static term}} + i  \underbrace{     \mu\int_0^t C^N_{II}(k,\tau)\mathcal{F}[\kappa^t(x_1)\Delta \dot{u}^t(x_1,t-\tau)] d\tau}_{\text{Dynamic term}},
\end{split}
\end{equation}
where $i$ is the complex number and $\mathcal{F}$ is the Fourier transform. 
The kernel has the expression:
\begin{equation}
\begin{split}
C^N_{II}(k,t) &=\left( \left(\frac{c_p^2}{c_s^2}-2\right)^2\frac{c_s^2}{2c_p^2}-\frac{c_s^2}{c_p^2}- c_s^2 k^2 t^2  \right) W(c_p k t ) + \left(1+c_s^2 k^2 t^2 -\frac{c_p^2}{2c_s^2} \right) W(c_s k t ) \\
&+ \left(\frac{c_s^2}{c_p^2}-\left(\frac{c_p^2}{c_s^2}-2\right)^2\frac{c_s^2}{2c_p^2}\right) J_1(c_p k t ) +\left(\frac{c_p^2}{2c_s^2}-1\right)J_1(c_s k t ) \\
&+ \frac{c_s^2}{c_p} k t  J_0(c_p k t )- c_s k t  J_0(c_s k t ).
\end{split}
\end{equation}
We note that this kernel is a pure imaginary term, the is because the kernel for normal traction is an odd function of position (as opposed to the kernel for shear traction that is an even function of position, hence, in wavenumber domain, the kernel for shear traction is a pure real). The static and dynamic terms perfectly cancel each other at $t=0$, as observed for shear traction.

%
%

%
%

\section{Model and algorithm}
We use a classic rate and state friction law \citep{dieterich1979a,ruina1983} with an aging law that is coupled with our previously derived sBIEM to perform a fully dynamic earthquake cycle simulation. The algorithm is based mainly on the methodology of \citet{lapusta2000} and \citet{lapusta2009a}. The wavenumber-dependent truncation of the integral is the one described in \citet{lapusta2009a} because we are inducing in-plane pure shear (mode II). The main difference is the adaptive time-stepping algorithm that we employ during the calculation, which is inspired by a Runge-Kutta method. 

\subsection{Normal and shear traction calculations}

\begin{figure}[H]
\centering
\includegraphics[width=\textwidth]{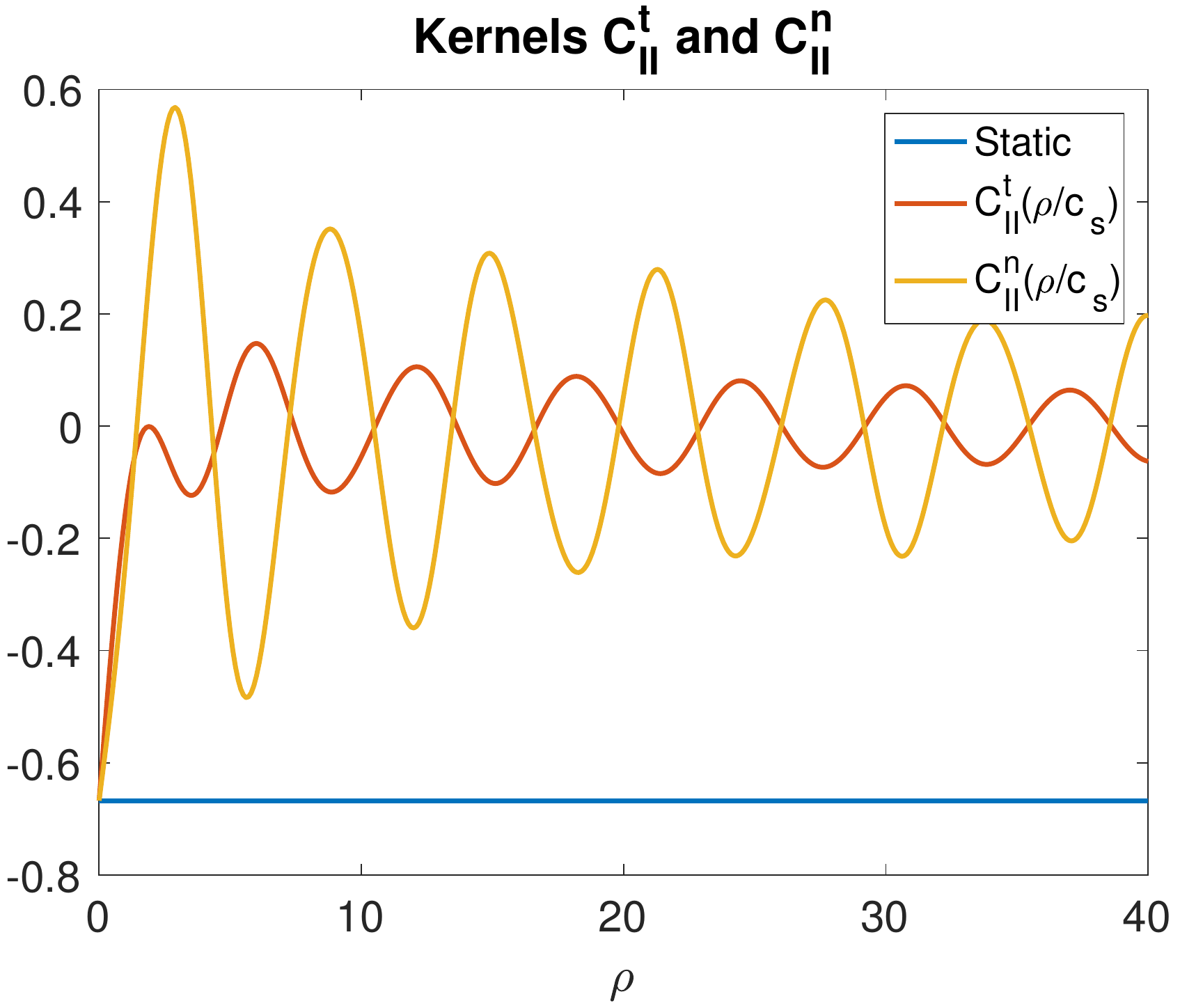}
\caption{Kernel representation for the shear and normal tractions. Please note that the kernel perfectly cancels the effect of the static term at $\rho=0$. }
\label{kernels}
\end{figure}

The normal and shear traction calculations are done in a similar manner to those in \citet{lapusta2009a}:
\begin{equation}
\begin{split}
\tau^{\text{el}}(k,t) &=  -\mu \left(1-\frac{c_s^2}{c_p^2} \right) |k|  \Delta u^t(k,t) - \mu |k| \int_0^{T_w} C^T_{II}(k,\tau) \Delta \dot{u}^t(k,t-\tau)d\tau, \\
\sigma_n^{\text{el}}(k,t) &= i \mu \left(1-\frac{c_s^2}{c_p^2} \right) \mathcal{F}[\kappa^t(x_1)\Delta u^t(x_1,t)] +   i    \mu  \int_0^{T_w} C^N_{II}(k,\tau)\mathcal{F}[\kappa^t(x_1)\Delta \dot{u}^t(x_1,t-\tau)] d\tau,
\end{split}
\end{equation}

where $T_w$ represents the time window over which the wavenumber-dependent integration is performed. Both the $C^T_{II}$ and $C^N_{II}$ kernels converge to zero as the argument goes to infinity (fig. \ref{kernels}). This kernel argument for $C^{T}_{II}$ and $C^{N}_{II}$ involves the multiplication of the wavenumber by the time ($k\tau$), such that the argument is large for high wavenumbers, even when a small time $\tau$ is considered. This is why we chose a cut-off wavenumber that was similar to \citet{lapusta2009a}:

\begin{equation}
T_w(k)= \left\{
	\begin{array}{ll}
		\frac{\eta \Omega }{c_s}  &  k\le k_c \\
		\frac{\eta \Omega}{ c_s }\frac{k_c}{k} & k\ge k_c
	\end{array}
\right.
\end{equation}
where $\eta$ is the truncation parameter (set to one in this study), $\Omega$ is the total domain size used in the sBIEM.
Note that the wavenumber-dependent truncation used in \citet{lapusta2000} is not efficient due to the slower convergence of the mode II kernel for the normal and shear tractions compared to the mode III kernel for shear traction \citep{lapusta2009a}. The amplitude for normal traction is around twice the amplitude for shear traction (see fig. \ref{kernels}), such that this approximation may require a higher $k_c$ value than that in the flat-fault case. One way to overcome this could be to introduce two different cutting planes, one for normal traction and one for shear traction. However, we did not think it was essential to incorporate this additional implementation in the code at this stage and we chose a single $k_c=\frac{200}{\eta \Omega}$.

\subsection{Time-adaptive solver}
\begin{figure}[H]
\centering
\includegraphics[width=\textwidth]{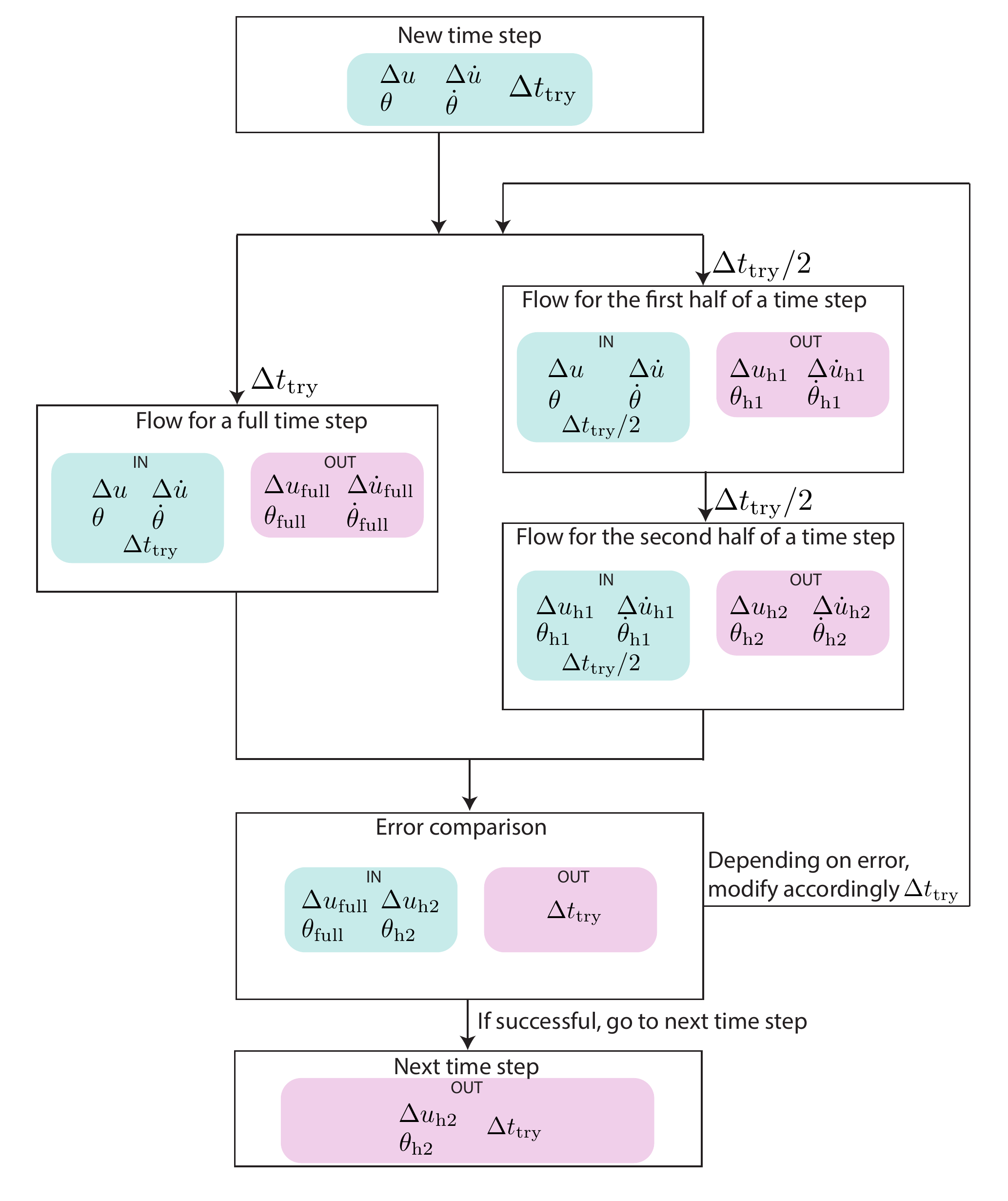}
\caption{Flow diagram of the adaptive time-step method that was used in this study, which was inspired by the fourth-order Runge-Kutta method (\cite{press1992}).}
\label{time_step} 
\end{figure}
The time-adaptive algorithm is different from the \citet{lapusta2000} approach and is inspired by the classic Runge-Kutta fourth-order method \citep{press1992}. Assuming that we want to employ a $dt$-second time step, we first do the classic flow for a full time step $dt$, and compute two smaller, consecutive half time steps $dt/2$ in parallel (fig. \ref{time_step}). The result using the two half time steps is generally more accurate than that for the full time step. We note that the second-order solutions for a full time step and two consecutive half time steps are $u_1^*$ and $u_2^*$, respectively. The exact solution at the next time step for the full time step $u_1(t+dt)$ and the two consecutive half time steps $u_2(t+dt)$ can be written as: 
\begin{equation}
\begin{split}
u_1(t+dt) &= \underbrace{u_1^*}_{\text{Second order solution}} +\underbrace{ \phi dt^3}_{\text{lowest order of error}}+ \underbrace{o(dt^4)}_{\text{Higher orders}} \\
and \\
u_2(t+dt) &= \underbrace{u_2^*}_{\text{Second order solution}} + \underbrace{ 2 \phi \left(\frac{dt}{2}\right)^3}_{\text{lowest order of error}}+ \underbrace{o(dt^4)}_{\text{Higher orders}},
\end{split}
\end{equation}
where $\phi$ is a constant over the time step \citep{press1992}. If we compare the two values in eq. 19, it is possible to better understand the error:
\begin{equation}
\begin{split}
\epsilon = u_1(t+dt) -u_2(t+dt)&= \frac{3}{4} \phi dt^3+ o(dt^4),
\end{split}
\end{equation}
which is a measure of the error at each time step. If we now want a given error $\epsilon_0$ at each time step, then we have to adapt the time step $dt_0$ to:
\begin{equation}
\begin{split}
dt_0 =  \left(\frac{\epsilon_0}{\epsilon} \right)^{\frac{1}{3}}dt.
\end{split}
\end{equation}
A given time step is considered successful if the error is smaller than the desired value. However, we adapt the error according to the previous equation, and consider a safety factor of $0.9$:
\begin{equation}
\begin{split}
dt_0 = 0.9  \left(\frac{\epsilon_0}{\epsilon} \right)^{\frac{1}{3}}dt.
\end{split}
\end{equation}
In the event that something went wrong during a given time step (e.g., the state became negative, Newton--Raphson did not converge, the normal traction became negative), then the next time step is simply divided by 2: 
\begin{equation}
\begin{split}
dt_0 = dt/2.
\end{split}
\end{equation}

We also decided that each time step should not be smaller than a given value:
\begin{equation}
dt_0 =\beta_{\text{min}} \Delta s/c_s,
\end{equation}
where $\beta_{\text{min}}$ is a coefficient that is smaller than one (same parameter as in \citet{lapusta2000}). $\beta_{\text{min}}$ is set to $0.25$ in this study. In the case the time step is already at the minimum value $dt =\beta_{\text{min}} \Delta s/c_s$, and the error made in the time step is greater than the set minimum error $\epsilon>\epsilon_0$, we decide to no not take into account the error and continue to the next time step. If another problem occurs (e.g., non-convergence of the Newton--Raphson algorithm) while the time step is already at the minimum, then the calculation stops.

\subsection{Time step}
The variables (slip: $\Delta u^t$, slip rate: $V$, state: $\theta$, and state rate: $\dot{\theta}$) are updated after the completion of a time step, following the methodology of \citet{lapusta2000}. The only modifications are the estimation of the traction change, which incorporates the normal traction variation, and the associated change in the momentum balance (eq. \eqref{momentum}). This scheme supposedly possesses second-order accuracy in time ($dt$) \citep{lapusta2000}. We note that the first-order estimates possess a 1 subscript ($\Delta u^t_1$, $V_1$, $\theta_1$ and $\dot{\theta}_1$) and the second-order estimates possess a 2 subscript ($\Delta u^t_2$, $V_2$, $\theta_2$ and $\dot{\theta}_2$). The 0 subscript represents the values of $\Delta u^t$, $V$, $\theta$, and $\dot{\theta}$ that are previously known at the beginning of the time step. 
\paragraph{(1) Slip and state estimation}
A first estimation of the slip and state at time $t+dt$ is obtained by using the velocity $V_0$ and state rate $\dot{\theta}_0$ at time $t$:
\begin{equation}
\begin{split}
\Delta u^t_1&= \Delta u^t_0 + V_0 dt \\
\theta_1&= \theta_0 + \dot{\theta}_0dt.
\end{split}
\end{equation}
\paragraph{(2) Traction estimation}
The traction is estimated by assuming that the velocity is constant over the time step. Our previous approximations of $\Delta u_1$, and $\theta_1$ allow us to calculate:
\begin{equation}
\begin{split}
\tau^{\text{el}}(k,t) &= -\mu \left(1-\frac{c_s^2}{c_p^2} \right) |k| \Delta u^t_1  \\
&-\mu |k|  \int_{dt}^{T_w} C^T_{II}(k,\tau) \Delta \dot{u}^t(k,t-\tau)d\tau \\
&- \Delta \dot{u}^t_1(k,t)\mu |k|  \int_{0}^{dt} C^T_{II}(k,\tau)d\tau, \\
\sigma_n^{\text{el}}(k,t) &= i \mu \left(1-\frac{c_s^2}{c_p^2} \right)  \mathcal{F}\left[  \kappa^t   \Delta \dot{u}^t_1 \right]  \\
&+i \mu  \int_{dt}^{T_w} C^N_{II}(k,\tau) \mathcal{F}\left[  \kappa^t   \Delta \dot{u}^t \right] d\tau \\
&+ \mathcal{F}\left[  \kappa^t   \Delta \dot{u}^t_1 \right] i \mu  \int_{0}^{dt} C^N_{II}(k,\tau)d\tau.
\end{split}
\end{equation}

The integrals are calculated using a midpoint method.

\paragraph{(3) Slip velocity and state rate estimation}
We solve for the equilibrium using the Newton--Raphson method, and the previous normal and shear traction estimates as follows:
\begin{equation}
\tau^{\text{load}}+\tau^{el} =( \sigma_n^{\text{load}}+\sigma_n^{\text{el}})\left(f_0+a \log \left(\frac{V_1}{V_0}\right)+b\log \left(\frac{\theta V_0}{D_c}\right)\right) ,
\label{momentum}
\end{equation}
where the velocity $V_1$ is the unknown variable, and $\sigma_n^{\text{load}}$ and $\tau^{load}$ are the normal and shear traction loads, respectively.
It is possible to simply update the state rate $\dot{\theta}$ once the velocity is updated using the state evolution law (aging law here) after solving for $V_1$ via eq. 27:
\begin{equation}
\dot{\theta}_1 =1-\frac{\theta_1 V_1}{D_c}.
\end{equation}
\paragraph{Reiteration of steps (1)--(3)}
We repeat steps (1)--(3) using the following values for the initial slip rate $V_0$ and state rate $\dot{\theta}_0$:
\begin{equation}
\begin{split}
V_0 &= \frac{V_0+V_1}{2} \\ 
\dot{\theta}_0 &= \frac{\dot{\theta}_0+\dot{\theta}_1}{2}.
\end{split}
\end{equation}
We then obtain accurate second-order values for the slip $\Delta u^t_2$, slip velocity $V_2$, state $\theta_2$, and state rate $\dot{\theta}_2$.
\paragraph{Update the final values}
We finally update the new values at $t+dt$, whereby all of the variables are their second-order estimates: $\Delta u^t(t+dt) = \Delta u^t_2$, $V(t+dt) = V_2$, $\theta(t+dt) = \theta_2$, and $\dot{\theta}(t+dt) = \dot{\theta}_2$. We also save the history of $V$ in the spectral domain for $t \in [t,t+dt]$ as $V(t) =\frac{V+V_1}{2}$, following \citet{lapusta2000}.


\section{Comparison with fully dynamic and quasi-dynamic space--time boundary element method}
We decided to test the algorithm against a fully dynamic (fdBIEM) and quasi-dynamic (qdBIEM) BIEMs in the space--time domain. The fdBIEM does not employ any approximations and is used as the reference here. The qdBIEM makes no approximation on the fault geometry but does not account for the wave-propagation in the medium. All of the parameters used in this study are listed in Table \ref{parameters_bend}. The periodic length of the fault in the sBIEM is four times the length of the fault in the qdBIEM and fdBIEM because the sBIEM requires an infinite, periodic fault. A total length of $1.5 \times L$ is constrained with a null slip rate on each side of the fault.

\begin{table*}\centering

\begin{tabular}{@{}lcc@{}}\toprule
Name & symbol & Value \\ \midrule
Reference friction coefficient &$f_0$ & $0.6$ \\
Reference velocity&$V_0$ & $10^{-9}$ m/s \\
Critical slip distance &$D_c$ & $1$ cm \\
Rate and state parameter&$a$ & $0.012$  \\ 
Rate and state parameter&$b$ & $0.015$  \\ \midrule
Initial normal stress & $\sigma_n$ & $100$ MPa \\
Shear modulus & $\mu $ & $40$ GPa \\
Shear velocity &$c_s$&  $3464$ m/s \\
Dilatational velocity &$c_p$  &  $6000$ m/s \\ \midrule
Fault length & $L$ & $10.240$ km \\ 
Resultant nucleation lengthscale$^1$ & $L_{\infty}$ & $5.659$ km \\ 
Resultant nucleation lengthscale$^2$ & $L_{b}$ & $355$ m \\
Discretization length& $ds$ & $10$ m \\ \midrule
\end{tabular}
\label{parameters_bend}
\caption{Parameters used in the simulation. $^1$ Two times the nucleation lengthscale obtained via \cite{rubin2005}. $^2$ The nucleation lengthscale obtained via \cite{dieterich1992}.}

\end{table*}

\subsection{The seamount case}
We chose a Gaussian geometry to test the effect of a seamount:
\begin{equation}
y_2 = A \exp(-(y_1-7)^2) ,
\end{equation}
where $y_1$ and $y_2$ are given in $km$.
We ran simulations for three different amplitude values $A \in \{0.03, 0.1, 0.3 \}$. 
\subsubsection{Results}
\begin{figure}[H] 
\centering
\includegraphics[width=\textwidth]{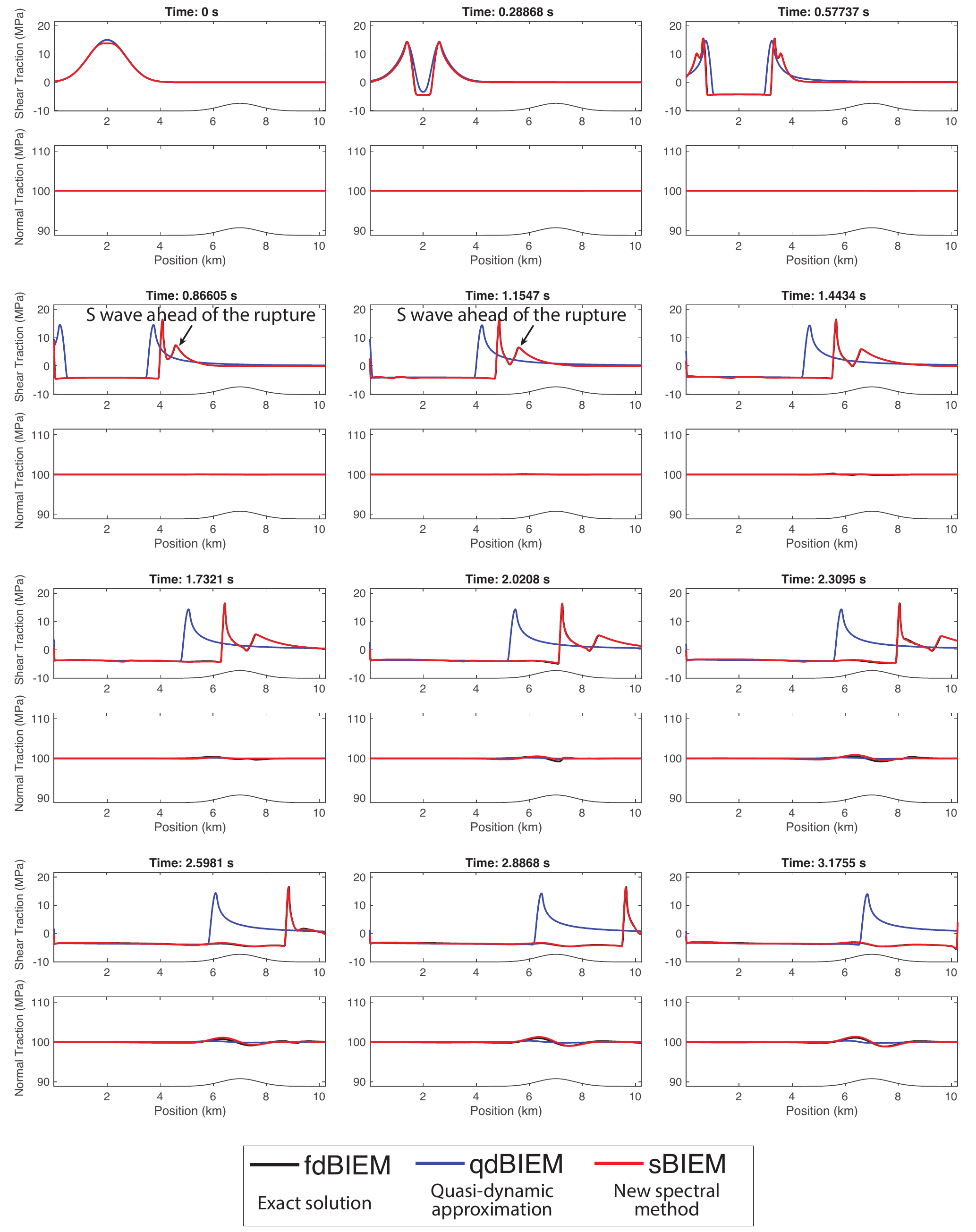}
\caption{Comparison of the normal and shear traction evolution for a seamount fault geometry obtained via a quasi-dynamic simulation using space--time BIEM (qdBIEM), a fully dynamic simulation using space--time BIEM (fdBIEM), and the method newly developed here; i.e., a fully dynamic simulation using spectral time BIEM (sBIEM). The amplitude of the seamount here is $A=30$ m.}
\label{tmp9} 
\end{figure}

\begin{figure}[H]
\centering
\includegraphics[width=\textwidth]{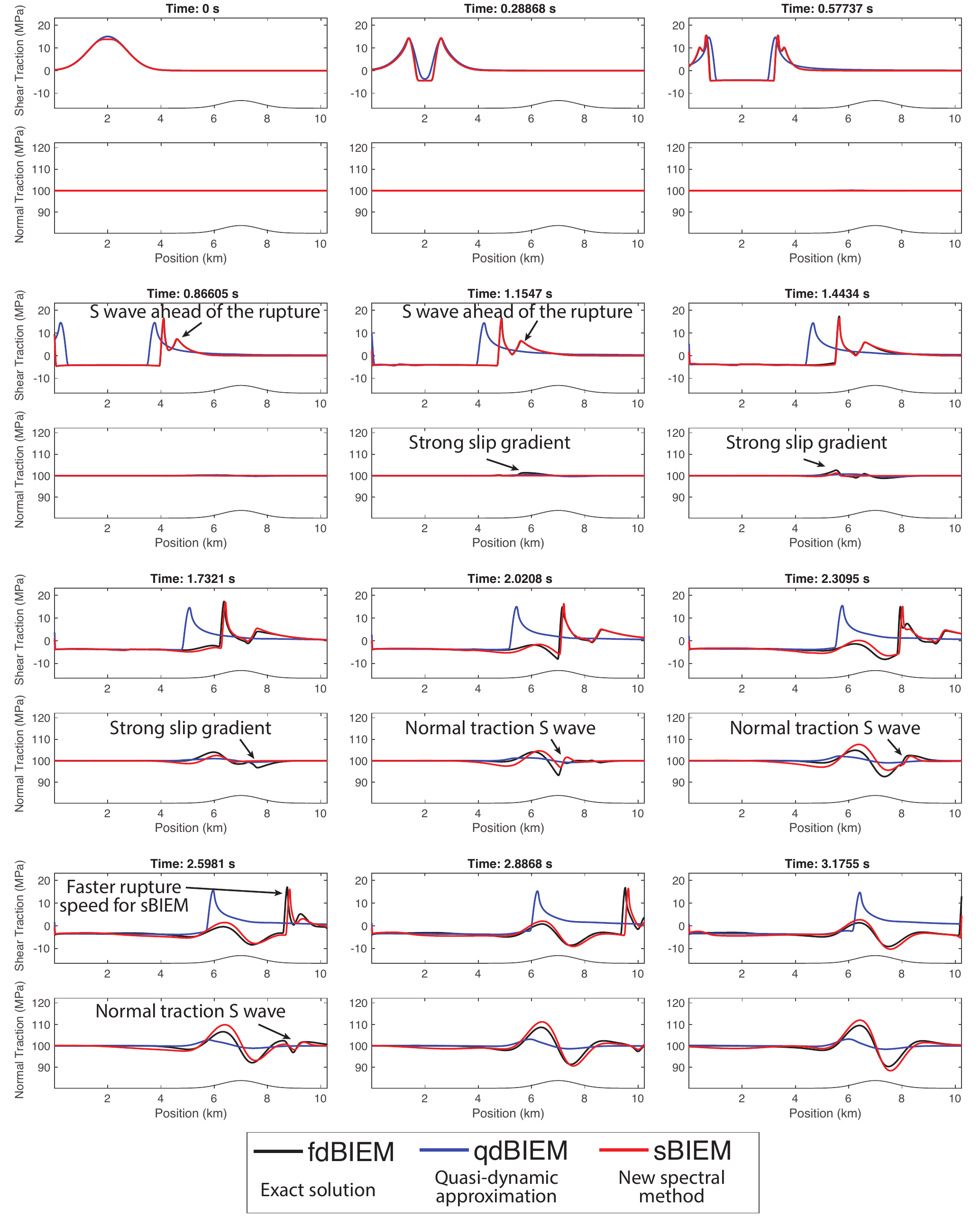}
\caption{Comparison of the normal and shear traction evolution for a seamount fault geometry obtained via a quasi-dynamic simulation using space--time BIEM (qdBIEM), a fully dynamic simulation using space--time BIEM (fdBIEM), and the method newly developed here; i.e., a fully dynamic simulation using spectral time BIEM (sBIEM). The amplitude of the seamount here is $A=300$ m.}
\label{tmp11} 
\end{figure}

\begin{figure}[H] 
\centering
\includegraphics[width=\textwidth]{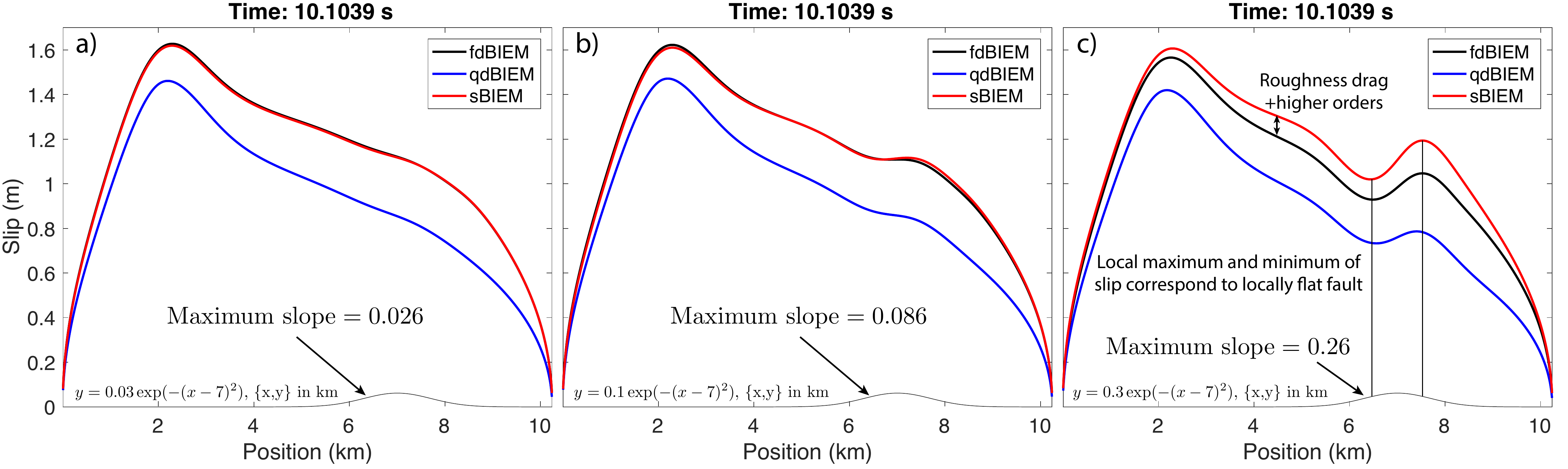}
\caption{a) Final slip distribution for a seamount fault geometry with amplitude $A = 30$ m. b) Final slip distribution for a seamount fault geometry with amplitude $A = 100$ m. c) Final slip distribution for a seamount fault geometry with amplitude $A = 300$ m.}
\label{seamount_slip} 
\end{figure}

Fig. \ref{tmp9} shows the evolution of the shear and normal tractions over time during the rupture for the qdBIEM, fdBIEM, and sBIEM models, under the assumption that the small-slope approximation is respected (case $A = 0.03$; maximum fault slope of $2.6\%$). The qdBIEM simulation is quite different from the fdBIEM and sBIEM simulations. This qdBIEM result is expected because there is no wave propagation and dynamic effects other than the radiation damping term in the model \citep{thomas2014c}. The rupture speed for the qdBIEM is significantly slower, and there is no S-wave at the rupture front. Furthermore, there are no reflected waves after the quasi-dynamic rupture ends. There is no visible difference between the fdBIEM and sBIEM models because the small-slope approximation is a valid assumption in this case. The results are still surprisingly good when we reach the point where the small-slope approximation is no longer a valid assumption (case $A=0.1$: maximum slope = $8.6\%$ (fig. \ref{tmp10}) and case $A=0.3$: maximum slope = $26\%$ (fig. \ref{tmp11})), although it is possible to observe some deviations. It can be seen that the sBIEM is over-estimating the maximum and minimum of the shear and normal tractions. Another noticeable difference is at the rupture tip when the rupture enters the seamount. This difference is probably due to the fact that the small-slope assumption neglects the slip gradient term ($1^{\text{st}}$ order term) in the normal traction calculation. In this simulation, since there is no prior slip on the fault, the situation where the slip gradient is much bigger than the fault curvature that multiplies the slip ($\frac{\partial}{\partial y^t} \Delta u^t \gg \kappa^t \Delta u^t $) arises. Hence, at the tip of the rupture where the slip gradients are high but the slip is small, the gradient term ($1\text{st}$ order term) in the calculation of normal traction is temporary bigger than the curvature that multiplies the slip term ($0^\text{th}$ order term). However, we expect this effect to be mitigated in multi-cycle simulations because the $\kappa^t \Delta u^t$ term will be larger (i.e., there will be prior slip before the rupture). 
The final difference is that the sBIEM yields a higher rupture speed. This enhanced rupture speed is clearly visible after the rupture passed the seamount. This difference may be explained by the fact that one of the higher-order terms is the roughness drag \citep{fang2013}, whereby there is an additional shear resistance that opposes movement on the two sides of the fault as soon as the fault becomes non-planar. This additional shear resistance consumes energy that would be otherwise available for the rupture to accelerate. However, this term is a higher-order term that we have neglected here (there is no additional shear resistance due to this seamount in the sBIEM). The final slip distributions for different seamount geometries are shown in fig. \ref{seamount_slip}. The slip distributions from the sBIEM and fdBIEM exhibit a high degree of agreement with each other until the slope reaches $8.6\%$ (fig. \ref{seamount_slip}-a and –b). The qdBIEM exhibits a smaller amount of slip on the fault. The sBIEM and fdBIEM no longer possess a high degree of agreement with each other for higher fault slopes (fig. \ref{seamount_slip}-c). The slip distribution from the sBIEM is significantly higher than that from the fdBIEM, although the general shapes of the slip distributions are similar. This is due to the fact that the sBIEM neglects the first-order term, which is the roughness drag \citep{fang2013}. We note that this is a useful way to show the effect of the roughness drag on a given slip distribution (fig. \ref{seamount_slip}-c).

\subsection{Rough fault}
We also test the rough-fault case. We generate a self-similar geometry using the Fourier transform method \citep{dunham2011b}. The fault profile has a spectral density of:
\begin{equation}
P(k)=(2\pi^3)\alpha^2 k^3,
\end{equation}
where $k$ is the wavenumber. We tried different amplitude-to-wavelength ratio $\alpha \in \{10^{-4},3 \times10^{-4},10^{-3}, 3\times10^{-3}\}$, and we set the minimum wavelength of the roughness to $20\Delta s =200 m$. The maximum slopes of the simulations are $\{ 0.0048, 0.014, 0.048, 0.14\}$, respectively.
\subsubsection{Results}
\begin{figure}[H] 
\centering
\includegraphics[width=\textwidth]{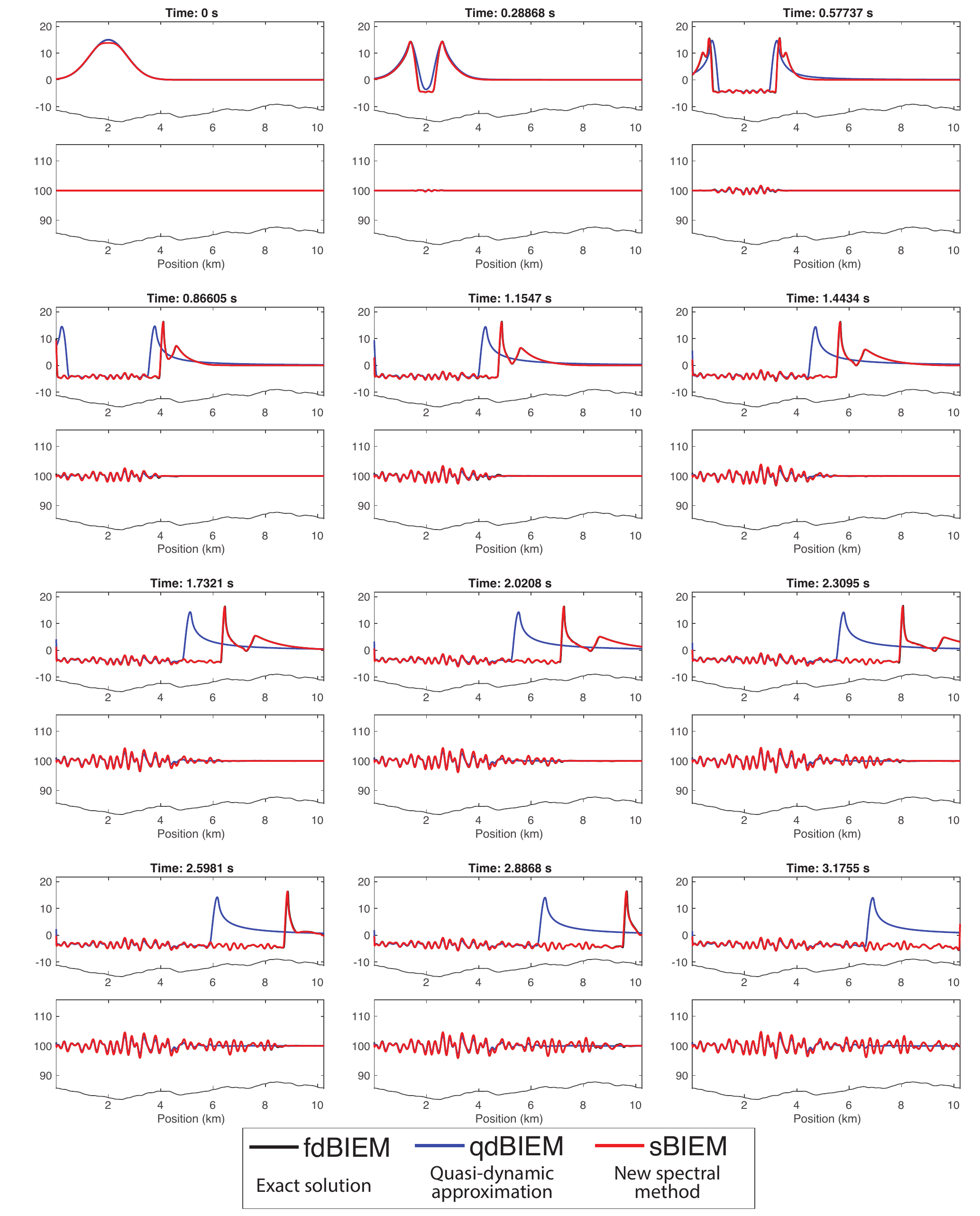}
\caption{Comparison of the normal and shear traction evolution of a rough-fault geometry obtained via a quasi-dynamic simulation using space--time BIEM (qdBIEM), a fully dynamic simulation using space--time BIEM (fdBIEM), and the method newly developed here; i.e., a fully dynamic simulation using spectral time BIEM (sBIEM). The amplitude-to-wavelength ratio is $\alpha=3 \times 10^{-4}$.}
\label{tmp15} 
\end{figure}

\begin{figure}[H] 
\centering
\includegraphics[width=\textwidth]{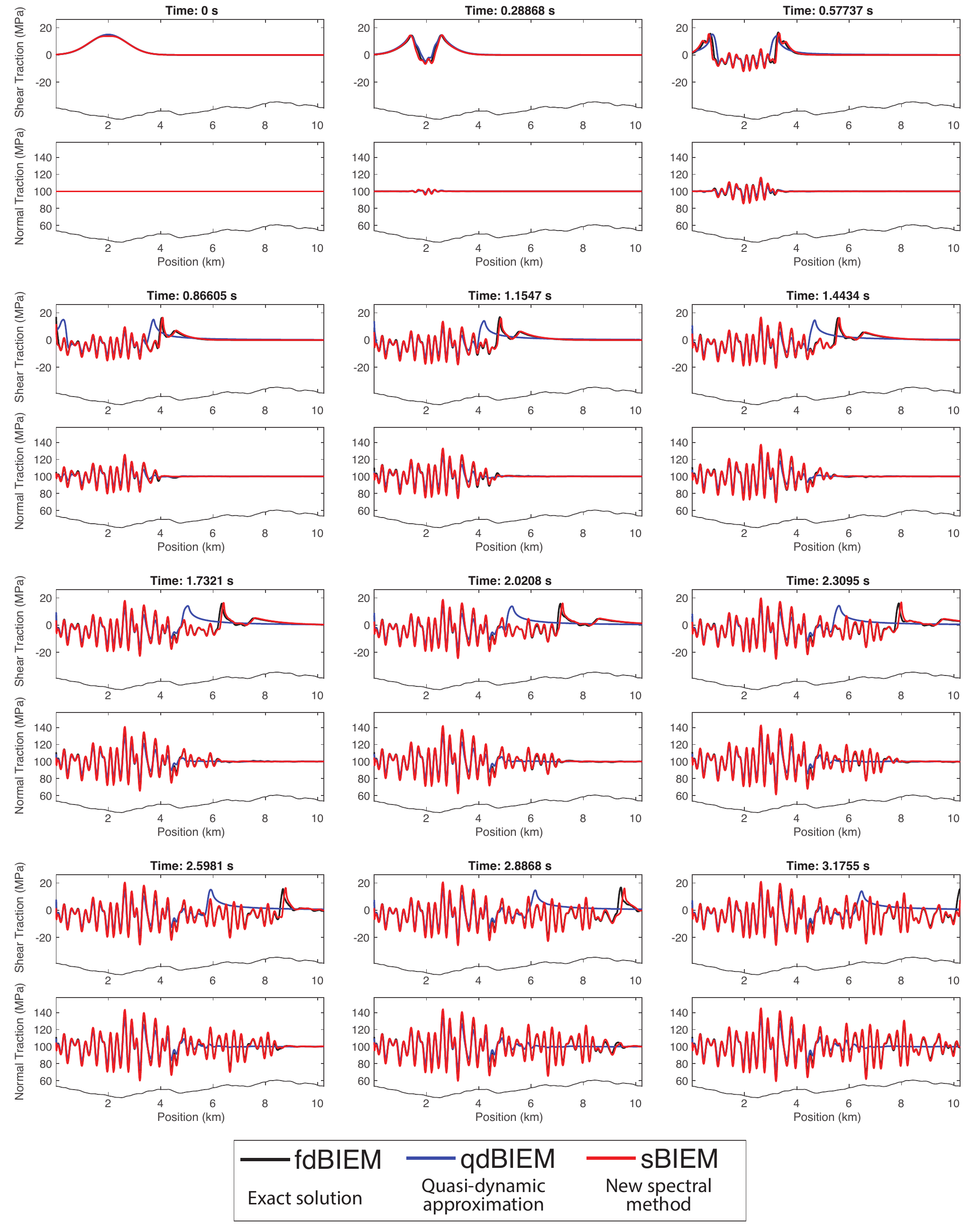}
\caption{Comparison of the normal and shear traction evolution for a rough-fault geometry via a quasi-dynamic simulation using space--time BIEM (qdBIEM), a fully dynamic simulation using space--time BIEM (fdBIEM), and the method newly developed here; i.e., a fully dynamic simulation using spectral time BIEM (sBIEM). The amplitude-to-wavelength ratio is $\alpha=3 \times 10^{-3}$.}
\label{tmp17}
\end{figure}

\begin{figure}[H] 
\centering
\includegraphics[width=\textwidth]{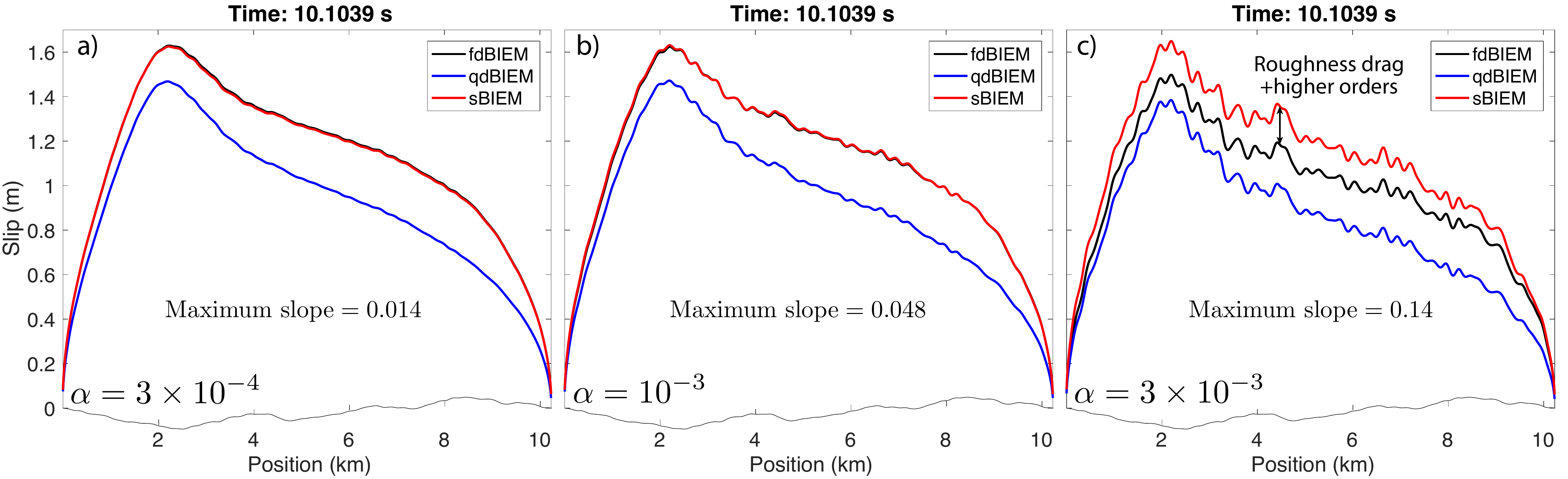}
\caption{a) Final slip distribution for a rough fault with an amplitude-to-wavelength ratio of $\alpha = 3 \times 10^{-4}$. b) Final slip distribution for a rough fault with an amplitude-to-wavelength ratio of $\alpha =10^{-3}$. c) Final slip distribution for a rough fault with an amplitude-to-wavelength ratio of $\alpha = 3 \times 10^{-3}$.}
\label{rough_slip}
\end{figure}

It is possible to see the normal and shear traction evolution along the fault for one mildly rough fault ($\alpha = 3\times10^{-4}$, maximum slope: $1.4\%$) where the small-slope approximation is valid, as shown in fig. \ref{tmp15}. The observations are very similar to the previous observations for the seamount. The sBIEM and fdBIEM results are in perfect agreement with each other, whereas the qdBIEM results exhibit a much lower rupture speed with no S waves ahead of the rupture. The case where the assumption of a small-slope approximation becomes invalid ($\alpha = 3\times10^{-3}$, maximum slope: $14\%$) is shown in fig. \ref{tmp17}. The sBIEM result is similar to the seamount geometry, whereby the sBIEM slightly over-estimates the rupture speed and the normal and shear tractions at the local extremas. The sBIEM and fdBIEM results are in agreement with each other as long as the small-slope approximation is valid (fig. \ref{rough_slip}-a and -b). However, the sBIEM begins to overestimate the slip compared to the true solution given by the fdBIEM as the small-slope approximation becomes invalid (fig. \ref{rough_slip}-c).

\section{Scaling of slip distribution versus fault geometry}
There is a simple scaling that links the scaled slip gradient with the curvature in this type of 2D in-plane simulation. This scaling was derived analytically by \citet{romanet2020}:
\begin{equation}
\frac{1}{f\Delta u^t}\frac{d}{dy^t}\Delta u^t = -\kappa^t.
\end{equation}
This scaling means that the slip gradient reaches a maximum (minimum) in the areas where the local curvature of the fault geometry is at a minimum (maximum). It also means that the local extrema of the slip correspond to the areas where the fault is locally flat ($\kappa^t=0$).

\begin{figure}
\centering
\includegraphics[width=1\textwidth]{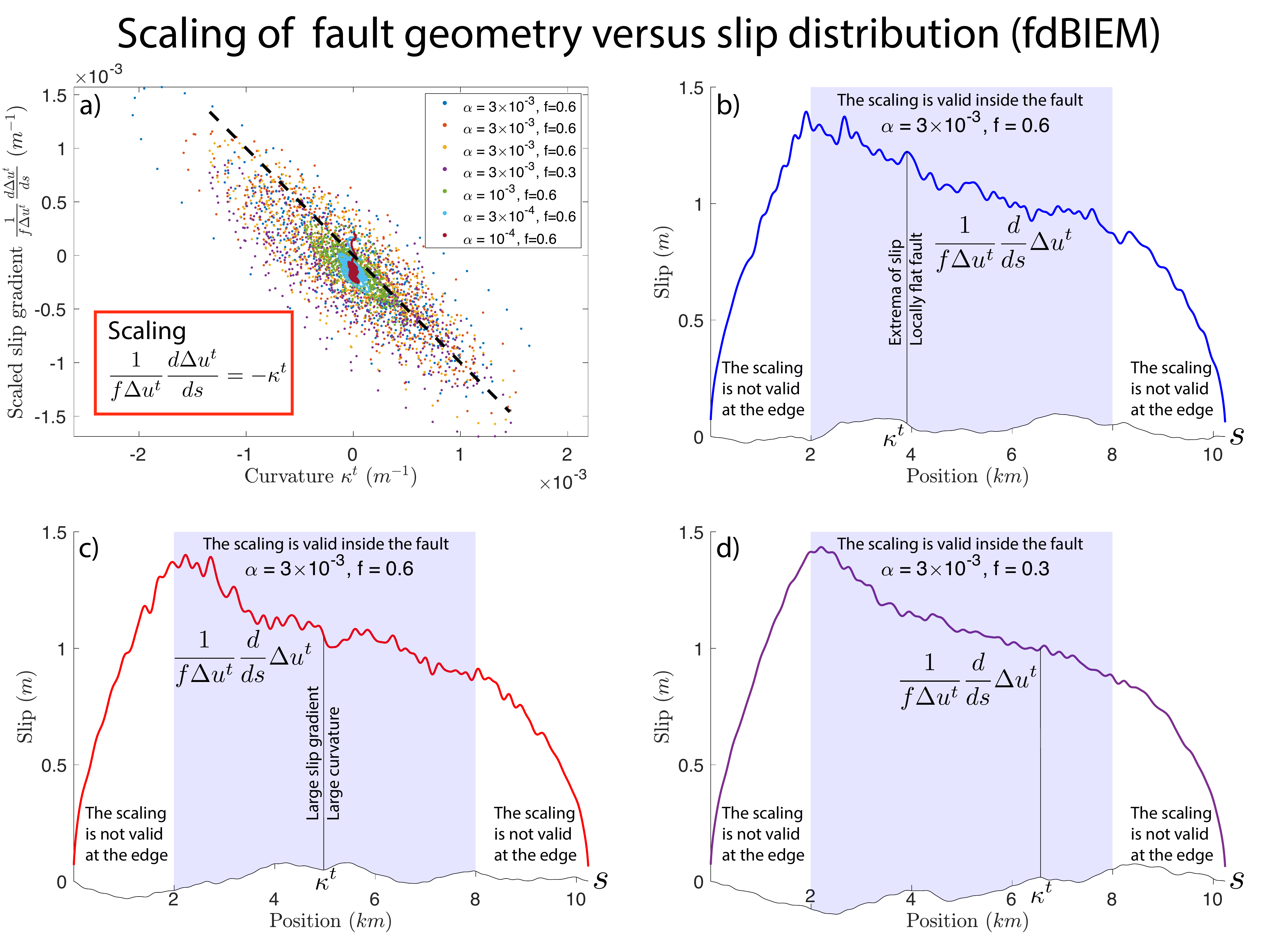}
\caption{a) Scaling of the slip distribution versus the curvature obtained using the classic space--time fully dynamic BIEM (fdBIEM). b) Slip distribution for a simulation with $\alpha = 3 \times 10^{-3}$ and $f=0.6$. c) Slip distribution for a simulation with $\alpha = 3 \times 10^{-3}$ and $f=0.6$. d) Slip distribution for a simulation with $\alpha = 3 \times 10^{-3}$ and $f=0.3$.}
\label{scaling}
\end{figure}

\begin{figure}
\centering
\includegraphics[width=1\textwidth]{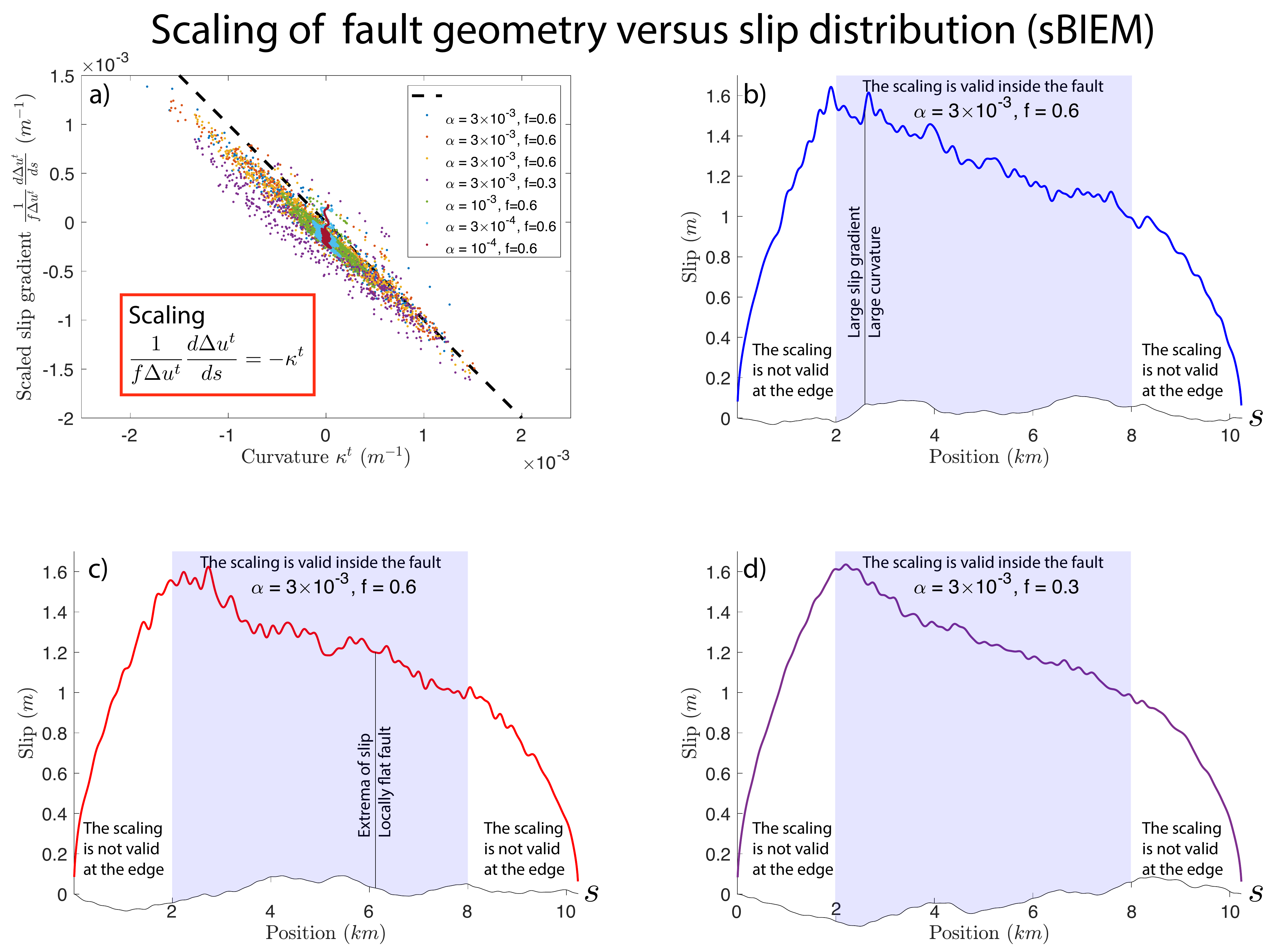}
\caption{a) Scaling of the slip distribution versus the curvature obtained using the newly developed spectral BIEM (sBIEM). b) Slip distribution for a simulation with $\alpha = 3 \times 10^{-3}$ and $f=0.6$. c) Slip distribution for a simulation with $\alpha = 3 \times 10^{-3}$ and $f=0.6$. d) Slip distribution for a simulation with $\alpha = 3 \times 10^{-3}$ and $f=0.3$.}
\label{scaling_sBIEM}
\end{figure}

We computed fully dynamic simulations using fdBIEM (fig. \ref{scaling}) and sBIEM (fig. \ref{scaling_sBIEM}) for different rough faults and different friction coefficients to test this scaling. 
The scaling results for several simulations with a friction coefficient of $f=0.6$ and  amplitude-to-wavelength ratio $\alpha \in \{10^{-4},3 \times10^{-4},10^{-3}, 3\times10^{-3}\}$ and for one simulation with $f=0.3$ and $\alpha = 3\times10^{-3}$ are shown in figs \ref{scaling} and \ref{scaling_sBIEM}, respectively. 
The scaling relationship (figs \ref{scaling}-a and \ref{scaling_sBIEM}-a) seems to be quite robust, although there is a small bias toward a negative gradient. This is due mainly to the fact that we triggered the earthquake artificially on the left part of the fault by adding an initial perturbation on the shear traction (it can be seen at time $t=0$ of the simulation on figs. \ref{tmp9},  \ref{tmp11},  \ref{tmp15}, and  \ref{tmp17}). This bias can be understood by looking at the slip distribution in figs \ref{scaling}-b--d and \ref{scaling_sBIEM}-b--d, where the slip in the middle of the fault has an overall decreasing derivative. If we focus on the scaling for the fdBIEM (fig. \ref{scaling}-a), it can be seen that the deviation from the theoretical scaling is slightly higher than that for the sBIEM (fig. \ref{scaling_sBIEM}-a). This is due to the fact that the theoretical scaling was also obtained by applying the small-slope approximation \citep{romanet2020}, as in the sBIEM. The fdBIEM simulation has higher-order terms that make its modeled slip distribution deviate from the theoretical scaling.
One interesting outcome of this scaling is that it is possible (at least theoretically) to invert for the geometry if we know the friction coefficient and slip distribution. 

\citet{bruhat2020} attempted to compare the observed slip distribution at the surface with a modeled slip distribution. Here we solve one part of the problem surrounding the scaling of the slip distribution and fault geometry for numerical studies. One important point is that the scaling is not valid close to the edge of the fault, as the slip gradients are high in this region. The edges of the fault should therefore be removed when conducting future analyses to obtain a scaling between the fault geometry and slip distribution.

\section{Seismic cycles}
\begin{figure}[H] 
\centering
\includegraphics[width=0.5\textwidth]{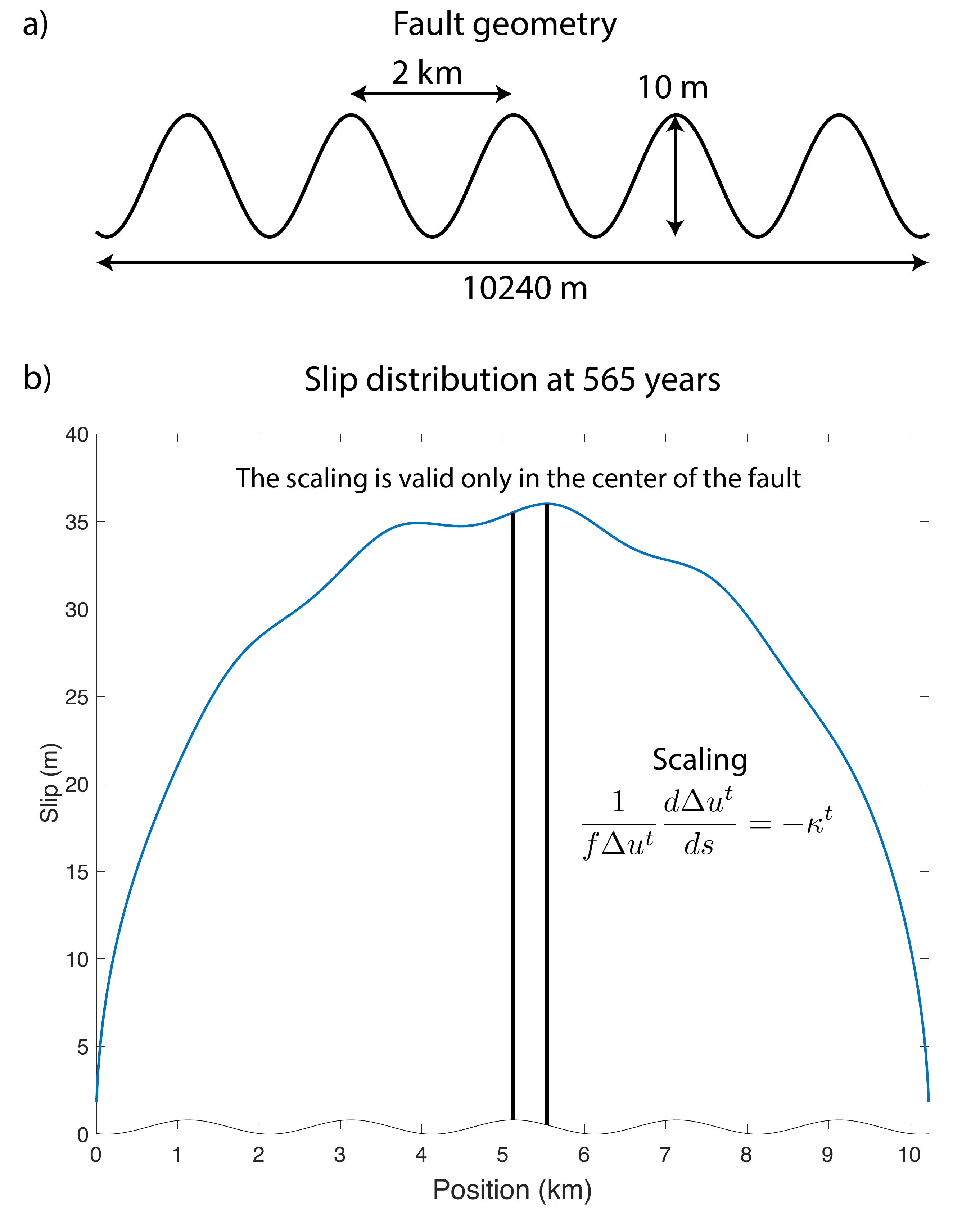}
\caption{Fully dynamic cycle simulation on a sinusoidal fault. a) Geometry of the fault. b) Final slip distribution at the end of the simulation.}
\label{geometry_cycle}
\end{figure}

\begin{figure}[H] 
\centering
\includegraphics[width=1\textwidth]{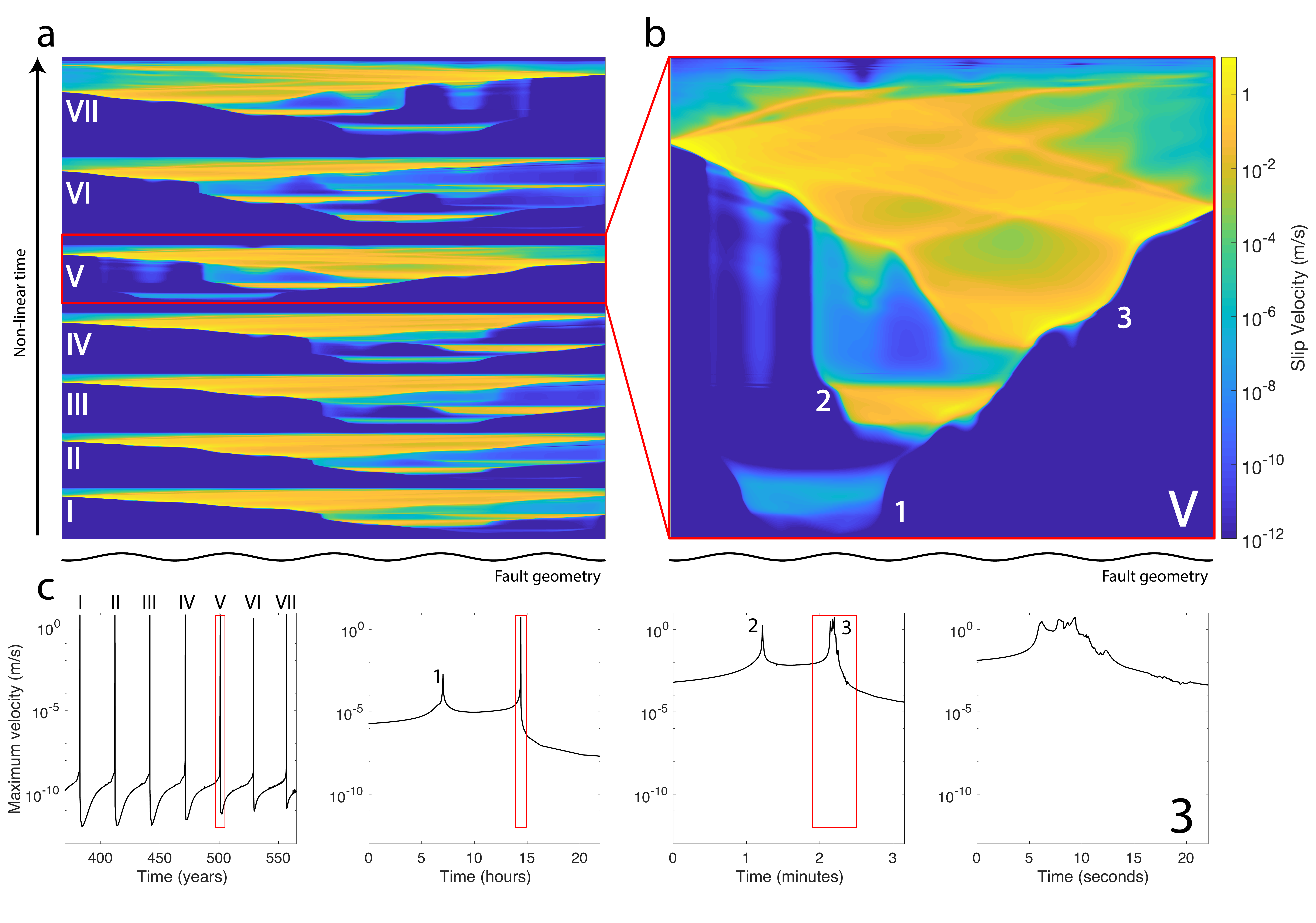}
\caption{Fully dynamic cycle simulation on a sinusoidal fault. a) Space--time evolution of the slip velocity on the fault. b) Closeup of the space--time evolution of the slip velocity for the $V^{\text{th}}$ event. c) Successive closeups of the $3^{\text{rd}}$ event, with the maximum velocity on the fault over time.}
\label{sliprate_cycle}
\end{figure}

We run one simulation of an earthquake cycle on a sinusoidal fault (fig. \ref{geometry_cycle}-a) using rate and state friction, and the sBIEM as a proof of concept. The model parameters are the same as those in Table \ref{parameters_bend}, except that the fault is loaded with a constant shear traction loading of $\tau^{\text{load}}= 0.01$ Pa/s along the fault. The loading is therefore completely uniform along the fault, even though the geometry is non-planar; we use this loading to avoid the effect of complex loading on the fault. The final slip distribution of the simulation is shown in fig. \ref{geometry_cycle}-b, where the scaling that links the fault geometry and slip distribution is respected. A particular example is in the center of the fault, where the maximum and minimum slip amounts correspond to a locally flat fault ($\kappa^t = 0$). The slip gradient is at a maximum when the fault has a high curvature ($\kappa^t\gg 1$). The slip rate evolution is shown in fig. \ref{sliprate_cycle}-a. The slip rate has a very complex evolution that produces both slow slip event (event 1 in fig. \ref{sliprate_cycle}-b) and foreshock (event 2 in fig. \ref{sliprate_cycle}-b). The cycle is largely periodic, such that the entire fault is ruptured periodically by a single earthquake (fig.  \ref{sliprate_cycle}-c). However, the events that occur prior to the mainshock show both spatial (their location on the fault changes between different main events) and temporal complexity, with an acceleration in the number of events before the mainshock (fig. \ref{sliprate_cycle}-c). This complex behavior of a foreshock sequence along a rough fault has already been noted by \citet{cattania2021}. One interesting point here is that the main event, which is preceded by foreshocks, becomes increasingly complex over time (\ref{sliprate_cycle}-a). This is because shear slip $\Delta u^t$ accumulates on the fault while the geometry is held constant, such that the curvature term $\kappa^t \Delta u^t$ continues to increase. This means that the normal traction variations along the fault continue to increase until the fault opens. Many methods have been employed to avoid this effect, such as loading with backslip \citep{heimisson2020}, placing a threshold on the normal traction \citep{cattania2021}, and including a visco-plastic effect \citep{dunham2011b}. Another possibility here would be to make the fault flatter by decreasing the curvature $\kappa^t$. This effect of a smoother fault with ongoing slip has already been observed on natural faults \citep{sagy2007,brodsky2011}.

\section{Discussion and conclusion}
We have shown that the small-slope approximation works quite well quantitatively if the fault slope is actually small. We are also able to qualitatively capture the stress variations along the fault due to its non-planarity for the case when the fault slope is not small. We emphasize that the fully dynamic results for the small-slope approximation are much closer to the full solution (space--time fdBIEM) than the qdBIEM results without the small-slope approximation. This implies that the zeroth-order effects of wave-mediated stress transfer are more important than the higher-order effect of fault non-planarity.

\subsection{Limits and future applications of this model}
There are several limitations of this method. One of the most constraining limitations is that this method is restricted to a homogeneous medium. This limitation will remain indefinitely, as we currently do not see a viable way to overcome this restriction. Another restriction is that this method is currently limited to a single fault. However, this does not appear to be a strong restriction because this can potentially be overcome via two different approaches. One is to employ the approach in \citet{barbot2021} and develop the spectral method for multiple parallel faults. The key drawbacks to this approach are that it is still limited to parallel faults and the development of a fully dynamic system may be complex. Another promising approach is to use the spectral method to model only the self-interaction of the faults on themselves, and account for the interaction between faults using the classic space--time BIEM. The key advantage of this approach is that other methods, such as H-matrices \citep{ohtani2011,bradley2014,sato2019}, can be used to accelerate the interaction between faults. This will make the H-matrix extremely efficient because we will get rid of the diagonal of the H-matrix (normally accounting for the self interaction) where compression (the rank reduction) is not possible.
Domain-based methods have also been recently employed for modeling the earthquake cycle with slow tectonic loading on a non-planar fault. Most of the methods employ a coupled static--dynamic method approach. Some methods implement only uni-directional coupling, with the quasi-dynamic model output used as the initial state for the dynamic model  \citep{galvez2019}. \citet{kaneko2011} coupled two spectral element methods, and \citep{liu2020} coupled two finite element methods. One recent study \citep{luo2020} adopted an adaptive dynamic relaxation technique to a finite element method, therefore allowing a fully dynamic simulation to be run in one unique finite element method framework.
These methods are more versatile than sBIEM because they allow for non-homogeneous media, free surfaces, complex fault geometries (without any assumptions), and multiple faults. However, these methods introduce absorbing surfaces in the simulations, and generally require much larger computation times than the sBIEM. Another advantage of the sBIEM is that it is a semi-analytical method (analytical in the form of an integral), therefore ensuring the convergence and accuracy of this method. With the exception of \citet{luo2020}, another advantage of the sBIEM over volumetric methods is the seamless transition between the interseismic and coseismic periods. Finally, the development of the sBIEM allows us to better understand the physics of complex fault geometries due to the analytical nature of this method. For example, we are able to analytically show the scaling between the fault geometry and slip distribution via the development of the sBIEM. We are also able to demonstrate that the main influence the fault geometry has on the earthquake cycle is the amount of normal traction on the fault itself. These analytical results could not have been obtained using only the volumetric method. They therefore highlight the need for diverse and complementary numerical methods to tackle a suite fault geometry and interaction scenarios in earthquake cycle simulations. 

\subsection{Development of a 3D fully dynamic sBIEM}
We are currently restricted to running the fully dynamic sBIEM in 2D. However, we are currently attempting to further develop this method in 3D using the space--time BIEM developed in \citet{romanet2020}. The 3D problem is much more complex: although the 2D fault geometry is only defined by one curvature, $\kappa^t$, we need to consider the four curvatures and two torsions associated with the slip field on the fault in 3D space \citep{romanet2020}. One reason for this additional complexity is that the slip direction is also allowed to change along the fault plane in 3D, which introduces the curvature due to changes in the slip direction. 

\subsection{Conclusion}
Here we have rigourously generalized the spectral boundary element method (sBIEM) for non-planar fault geometries. We have shown that the main effect of non-planar fault geometry in a fully dynamic 2D medium is to modify the normal traction along the fault. We have incorporated this generalized sBIEM into an existing methodology to model the fully dynamic earthquake cycle on a non-planar fault. Finally, we have tested our new method against the classic space--time BIEM to determine the limits of our method. We have demonstrated that this method agrees with the classic BIEM if the small-slope approximation is respected along the fault. The sBIEM continued to yield results that were still in good agreement with the true results once the small-slope approximation was no longer valid, although there was some quantitative difference between the modeled and true results (higher rupture speed, higher slip, and overestimation of the maxima and minima of the normal and shear tractions). One of the most important aspects of this method was the preservation of the scaling between the slip distribution and fault geometry. We hope that this method will provide a fast and convenient way to better understand the effect of the fault geometry on the earthquake cycle.

\section*{Author contribution statement}
P.R. wrote the manuscript, wrote the spectral code, and came up with the original idea of the paper. S.O. wrote the classic boundary integral element code. S.O. and P.R. both participated in producing and discussing the numerical results. Both authors have read and approved the manuscript.

\section*{Acknowledgement}
This work would not have been possible without the support of Satoshi Ide, Robert Viesca, Ryosuke Ando, Tatsuhiko Saito, and Raul Madariaga.

\section*{Data and Resources}
No data were used in this paper.



\clearpage
\bibliography{/Users/pierre/Dropbox/Public/CollectedPapers/MasterBibliography}
\bibliographystyle{./agufull08}

\newpage  
\noindent
ROMANET Pierre \\
Earthquake and Tsunami Research Division,\\
National  Research Institute for Earth Science and Disaster Resilience, \\
3-1, Tennodai, Tsukuba, Ibaraki, 305-0006, JAPAN \\
E-mail: romanet@bosai.go.jp
\\ \\
OZAWA So  \\
Department of Earth and Planetary Science, School of Sciences,  The University of Tokyo, \\
7-3-1 Hongo, Bunkyo-ku, Tokyo 113-0033, JAPAN \\
E-mail: sozawa@eps.s.u-tokyo.ac.jp

\newpage

\appendix
\numberwithin{equation}{section}
\counterwithin{figure}{section}
%
\section{Calculation of primitives}
\label{calculationT}
\subsection{Primitive of $T  J_{i_1}(aT)$}
We used the table by \citet{rosenheinrich2016} to calculate the Bessel function primitive. It can also be noted that the function $W$ is simply the opposite of the Bessel integral function of the first-order $J_{i_1}$ \citep{humbert1933}:
\begin{equation}
\begin{split}
 W(x) &= \int_x^{\infty}\frac{J_1(t)}{t}dt \\
 &= -J_{i_1}(x),
 \end{split}
\end{equation}
which leads to:
\begin{equation}
\begin{split}
\int_0^t x J_{i_1}(ax) dx&= \left[\frac{x^2}{2}J_{i_1}(ax)  \right]_0^t-a \int_0^t \frac{x^2}{2}\frac{J_1(ax)}{ax}dx \\
&= \frac{t^2}{2}J_{i_1}(at) - \frac{1}{2}\int_0^t x J_1(ax)dx\\
&= \frac{t^2}{2}J_{i_1}(at) -  \frac{1}{2} \int_0^{at} \frac{x'}{a^2}J_1(x')dx' \\
&= \frac{t^2}{2}J_{i_1}(at) -  \frac{1}{2a^2} \int_0^{at}x'J_1(x')dx' \\
&= -\frac{t^2}{2} (1-at J_0(at)+J_1(at)-\psi(at)) - \frac{1}{2a^2}  \psi(at)\\
\end{split}
\label{ji1}
\end{equation}
where $H_i$ is the Struve function. Substitution of eq. \eqref{psi} into eq. \eqref{ji1} yields:
\begin{equation}
\begin{split}
\int_0^t x W(ax) dx&= \frac{t^2}{2}W(at)+\frac{1}{2a^2}\psi(at).
\end{split}
\end{equation}

\subsection{Primitive of $J_0(aT)$}
\begin{equation}
\begin{split}
\int_0^t  J_{0}(ax) dx&=\frac{1}{a}\int_0^{at}  J_{0}(X) dX\\
&=t J_0(at)+\frac{\psi(at)}{a}
\end{split}
\end{equation}
\subsection{Primitive for shear}
\begin{equation}
C_{II}^T(k,x)= \frac{J_1(c_s k x)}{ x}+4c_s^2 k^2 x [W(c_p k x)-W(c_s k x)]-4\frac{c_s^2}{c_p} kJ_0(c_p k x )+3c_s k J_0(c_s k x)
\end{equation}

\begin{equation}
\begin{split}
\int C_{II}(t)dt&=(1-W(c_s kt)) \\
&+ 4c_s^2 k^2 \left(\frac{t^2}{2}W(c_p k t ) -\frac{t^2}{2}W(c_s k t)\right)+4c_s^2 k^2\left(\frac{1}{2c_p^2k^2}\psi(c_p k t) - \frac{1}{2c_s^2k^2}\psi(c_s k t)\right)  \\
   &  -4 \frac{c_s^2}{c_p}  k ( t J_0(c_p k t )+\frac{1}{c_p k } \psi(c_p k t))+3 c_s k t J_0(c_s k t)+3\psi(c_s k t)\\
      &=2\left(1-\frac{c_s^2}{c_p^2} \right) \\
    &+\left(2c_s^2 k^2t^2+2\frac{c_s^2}{c_p^2}\right)W(c_p k t)-\left(2c_s^2 k^2t^2+2\right)W(c_s k t) \\
    &-2\frac{c_s^2}{c_p^2}J_1(c_p kt)+J_1(c_s k t)\\
    &-2\frac{c_s^2}{c_p}ktJ_0(c_p k t)+2c_s k tJ_0(c_s k t)
 \end{split}
\end{equation}

\section{Details of the normal traction calculation}
\label{calculationN}

We can perform an integration by parts, such that the previous result becomes:
\begin{equation}
\begin{split}
&\sigma_n^{\text{el}}(\mathbf{x},t) = \\
&\int_{0}^{t}  \int _{\Sigma}  \left[  \frac{4c_s^2 \mu(t-\tau)}{2\pi(x_1-y_1)^3}\sqrt{(t-\tau)^2-\frac{(x_1-y_1)^2}{c_p^2}}+\frac{c_s^2\lambda^2(t-\tau)}{2\pi c_p^2(x_1-y_1)\sqrt{(t-\tau)^2-\frac{(x_1-y_1)^2}{c_p^2}}} \right]  \\
& \qquad \qquad  \left[  \kappa^t   \Delta \dot{u}^t \right]  \mathcal{H}\left( t-\tau-\frac{ | x_1-y_1|}{c_p} \right) d\Sigma d\tau  +\\
-&\int_{0}^{t}  \int _{\Sigma}  \left[  \frac{4c_s^2 \mu(t-\tau)}{2\pi(x_1-y_1)^3}\sqrt{(t-\tau)^2-\frac{(x_1-y_1)^2}{c_s^2}}+\frac{c_s^2(\lambda+2\mu)(t-\tau)}{2\pi c_s^2(x_1-y_1)\sqrt{(t-\tau)^2-\frac{(x_1-y_1)^2}{c_s^2}}}  \right]  \\
& \qquad \qquad  \left[  \kappa^t   \Delta \dot{u}^t \right]   \mathcal{H}\left( t-\tau-\frac{ | x_1-y_1|}{c_s} \right)d\Sigma d\tau,  
\end{split}
\end{equation}
where $\lambda$ is a Lam\'{e} parameter. 
We can prepare the previous equation for the Fourier transform (it is possible to develop the equation at this stage using Taylor expansion to obtain the static term):
\begin{equation}
\begin{split}
&\sigma_n^{\text{el}}(\mathbf{x},t) = \\
&\int_{0}^{t}  \int _{\Sigma}  \left[  \frac{4c_s^2 \mu}{2\pi c_p^3(t-\tau)}\frac{c_p^3(t-\tau)^3}{(x_1-y_1)^3}\sqrt{1-\frac{(x_1-y_1)^2}{c_p^2(t-\tau)^2}}+\frac{c_s^2\lambda^2}{2\pi\mu c_p^3(t-\tau)}\frac{c_p(t-\tau)}{(x_1-y_1)\sqrt{1-\frac{(x_1-y_1)^2}{c_p^2(t-\tau)^2}}} \right] \\
& \qquad \qquad  \left[  \kappa^t   \Delta \dot{u}^t \right]   \mathcal{H}\left( t-\tau-\frac{ | x_1-y_1|}{c_p} \right) d\Sigma d\tau  +\\
-&\int_{0}^{t}  \int _{\Sigma}  \left[  \frac{4 \mu}{2\pi c_s(t-\tau)}\frac{c_s^3(t-\tau)^3}{(x_1-y_1)^3}\sqrt{1-\frac{(x_1-y_1)^2}{c_s^2(t-\tau)^2}}+\frac{(\lambda+2\mu)}{2\pi c_s(t-\tau)}\frac{c_s(t-\tau)}{(x_1-y_1)\sqrt{1-\frac{(x_1-y_1)^2}{c_s^2(t-\tau)^2}}} \right] \\
& \qquad \qquad  \left[  \kappa^t   \Delta \dot{u}^t \right] \mathcal{H}\left( t-\tau-\frac{ | x_1-y_1|}{c_s} \right)   d\Sigma d\tau.  
\end{split}
\end{equation}

If we consider the two Fourier transforms:
\begin{equation}
\mathcal{F}\left[ \frac{\text{rect}(\frac{x}{2b})}{\frac{x}{b}\sqrt{1-\frac{x^2}{b^2}}}\right] =- i \pi b \left( 1+J_1(bk)-W(bk) \right)
\end{equation}

and

\begin{equation}
\begin{split}
\mathcal{F}\left[\frac{b^3}{x^3} \sqrt{1-\frac{x^2}{b^2}} \text{rect}(\frac{x}{2b})\right] &=  \frac{1}{2}i \pi b (bk J_0(bk) +J_1(bk) +1 +b^2k^2 -(1+b^2k^2)W(bk)),
\end{split}
\end{equation}

the properties of the Fourier transform:
\begin{equation}
\begin{split}
\mathcal{F}[f(x)] &=F(k)\\
\mathcal{F}[f(ax)]&= \frac{1}{|a|}F\left(\frac{k}{a}\right),
\end{split}
\end{equation}

and the relationship between the Lam\'{e} parameters:
\begin{equation}
\frac{\lambda^2}{\mu^2}=\left(\frac{c_p^2}{c_s^2}-2 \right)^2,
\end{equation}
and \\
\begin{equation}
\lambda+2\mu = \mu\frac{c_p^2}{c_s^2},
\end{equation}
it then becomes possible to rewrite $\sigma_n(\mathbf{x},t)$ in the wavenumber domain: 
\begin{equation}
\begin{split}
&\sigma_n^{\text{el}}(k,t) = \\
&\int_{0}^{t}   [\frac{4c_s^2 \mu}{2\pi c_p^3(t-\tau)}  ( \frac{1}{2}ic_p \pi(c_p k J_0(c_p k) +J_1(c_p k) +1 +c_p^2k^2 -(1+c_p^2k^2)W(c_p k))) \\
  &-\frac{c_s^2\lambda^2}{2\pi\mu c_p^3(t-\tau)}  i c_p \pi \left( 1+J_1(c_p k)-W(c_p k) \right)  ]  \\
& \qquad \qquad  \mathcal{F}\left[  \kappa^t   \Delta \dot{u}^t \right]   d\Sigma d\tau  \\
-&\int_{0}^{t} [  \frac{4 \mu}{2\pi c_s(t-\tau)}(  \frac{1}{2}ic_s \pi(c_s k J_0(c_s k) +J_1(c_s k) +1 +c_s^2k^2 -(1+c_s^2k^2)W(c_s k))
)\\ &-\frac{(\lambda+2\mu)}{2\pi c_s(t-\tau)} i c_s \pi\left( 1+J_1(c_s k)-W(c_s k) \right) ] \\
& \qquad \qquad  \mathcal{F}\left[  \kappa^t   \Delta \dot{u}^t \right]   d\Sigma d\tau.
\end{split}
\end{equation}

\section{Classic space--time BIEM}
\subsection{Fully dynamic BIEM}
The classic BIEM for the 2D plane strain problem was first developed by \citet{tada1997}. 
We used the discretized kernel given by eqs 4--6 in \citet{ando2007} and the second-order accuracy time stepping provided by \citet{noda2020}.
The computational code is available at github.com/sozawa94/fd2d.

\subsection{Quasi-dynamic BIEM}
The effect of the past slip rate is approximated via the radiation damping term \citep{rice1993}, which removes the convolution over time. Only the static kernel has to be convolved over space (eqs. 19-21 in \citet{ando2007}).
An adaptive time step is employed to control the error estimated by the difference between the fourth- and fifth-order solutions \citep{press1992}.
The computational code is available at github.com/sozawa94/hbi.

 \section{Additional figures}

 \begin{figure} 
\centering
\includegraphics[width=\textwidth]{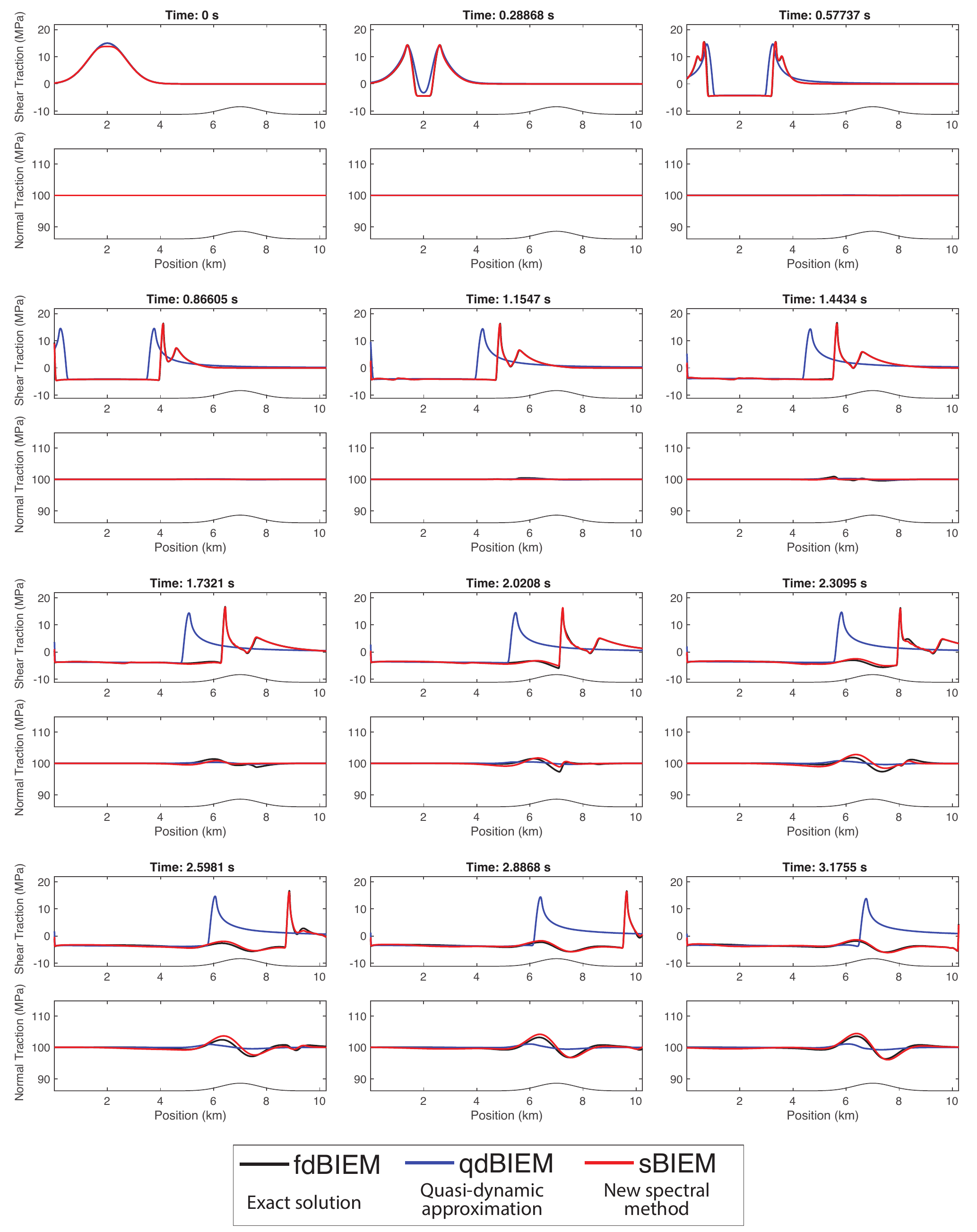}
\caption{Comparison of the normal and shear traction evolution for a seamount fault geometry obtained via a quasi-dynamic simulation using space--time BIEM (qdBIEM), a fully dynamic simulation using space--time BIEM (fdBIEM), and the method newly developed here; i.e., a fully dynamic simulation using spectral time BIEM (sBIEM). The amplitude of the seamount is $A=100$ m.}
\label{tmp10}
\end{figure}

 \begin{figure} 
\centering
\includegraphics[width=\textwidth]{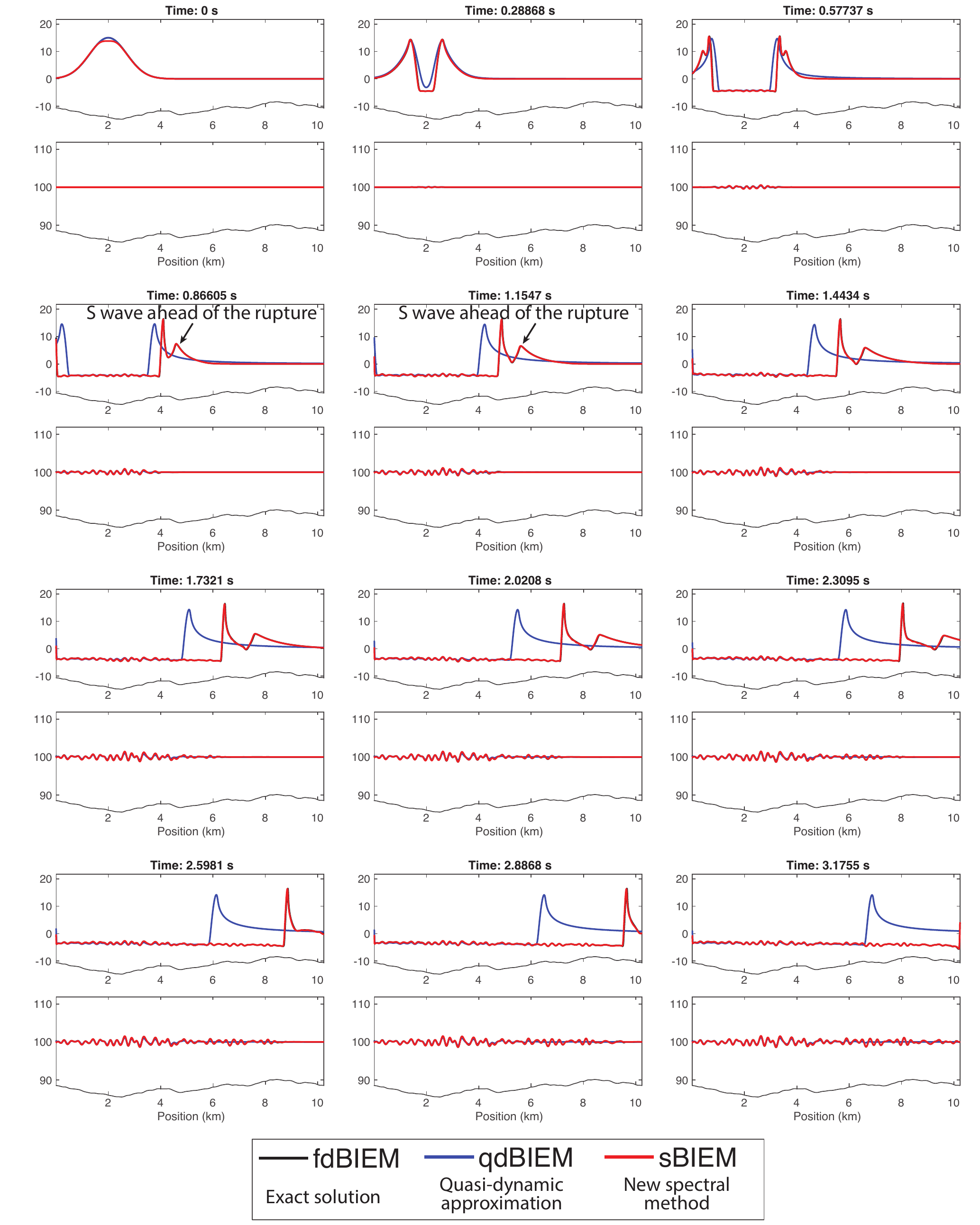}
\caption{Comparison of the normal and shear traction evolution for a rough fault geometry obtained via a quasi-dynamic simulation using space--time BIEM (qdBIEM), a fully dynamic simulation using space--time BIEM (fdBIEM), and the method newly developed here; i.e.,  a fully dynamic simulation using spectral time BIEM (sBIEM). The amplitude-to-wavelength ratio here is $\alpha=1 \times 10^{-4}$.}
\label{tmp14}
\end{figure}

 \begin{figure} 
\centering
\includegraphics[width=\textwidth]{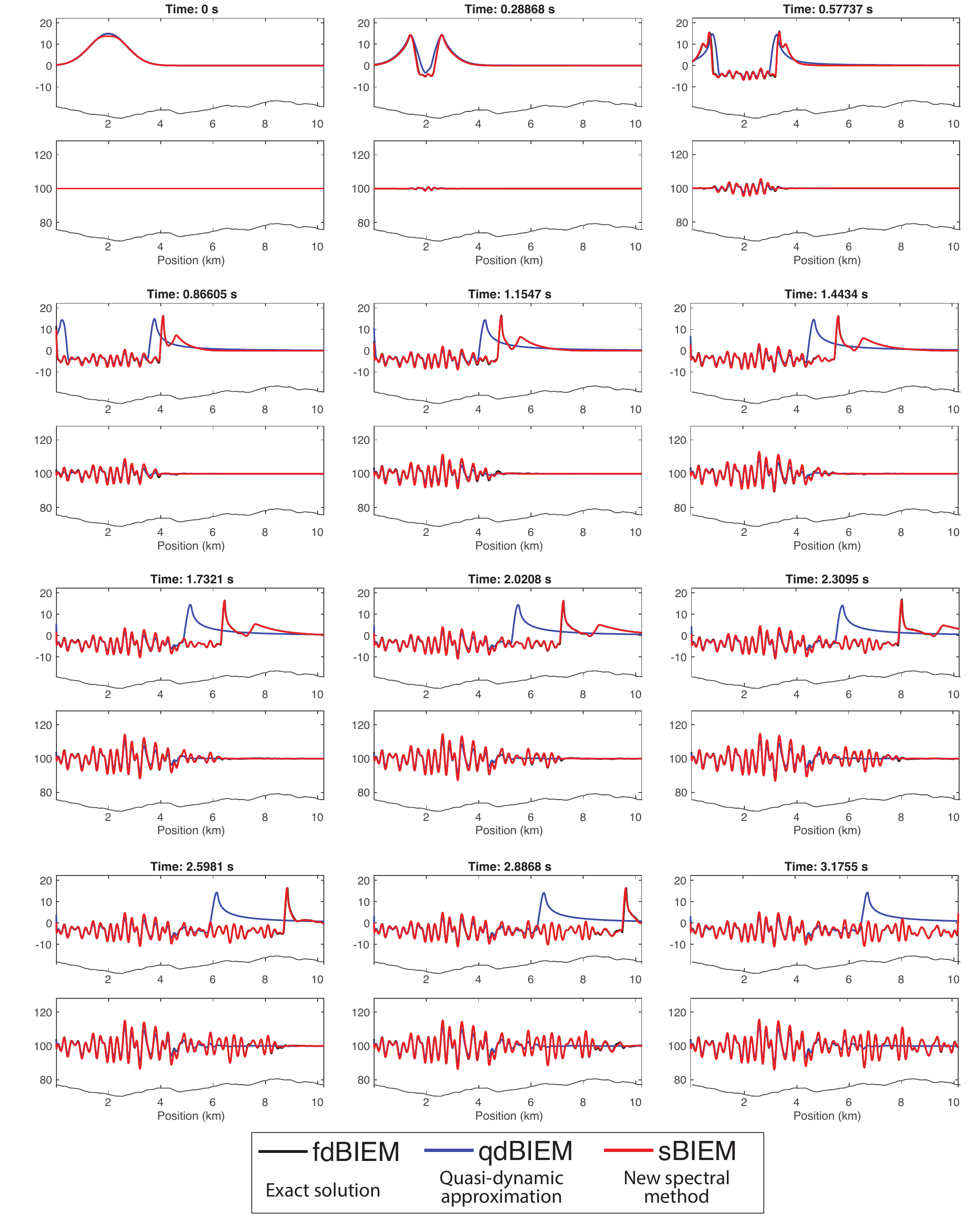}
\caption{Comparison of the normal and shear traction evolution for a rough fault geometry obtained via a quasi-dynamic simulation using space--time BIEM (qdBIEM), a fully dynamic simulation using space--time BIEM (fdBIEM), and the method newly developed here; i.e., a fully dynamic simulation using spectral time BIEM (sBIEM). The amplitude-to-wavelength ratio here is $\alpha=1 \times 10^{-3}$.}
\label{tmp16}
\end{figure}

\end{document}